\documentclass[useAMS,usenatbib]{mn2e}

\RequirePackage{natbib}
\usepackage{graphicx}
\usepackage{amsmath}
\usepackage{amsfonts}
\usepackage{amssymb}
\usepackage{breqn}
\usepackage{mathtools}

\title[Formation of massive seed black holes via collisions and accretion]{Formation of massive seed black holes via collisions and accretion}

\author[T.~Boekholt, D.~Schleicher, M.~Fellhauer, R.~Klessen, B.~Reinoso, A.~Stutz and L.~Haemmerl{\'e}]{T.~C.~N.~Boekholt$^{1,2}$\thanks{E-mail: boekholt@ua.pt (TB); dschleicher@astro-udec.cl (DRGS); mfellhauer@astro-udec.cl (MF), klessen@uni-heidelberg.de (RSK), breinoso@udec.cl (BR), astutz@astro-udec.cl (AMS) and \mbox{Lionel.Haemmerle@unige.ch (LH)}.}, D.~R.~G.~Schleicher$^{1}$\footnotemark[1], M.~Fellhauer$^{1}$\footnotemark[1], R.~S.~Klessen$^{3,4}$\footnotemark[1], 
\newauthor B.~Reinoso$^{1}$\footnotemark[1], A.~M.~Stutz$^{1}$\footnotemark[1], and L.~Haemmerl{\'e}$^{5}$\footnotemark[1] \\
$^{1}$Departamento de Astronom\'ia, Facultad Ciencias F\'isicas y Matem\'aticas, Universidad de Concepci\'on, Av. Esteban Iturra s/n \\Barrio Universitario, Casilla 160, Concepci\'on, Chile.\\
$^{2}$CIDMA, Departamento de F\' isica, Universidade de Aveiro, Campus de Santiago, 3810-193 Aveiro, Portugal.\\ 
$^{3}$Universit\" at Heidelberg, Zentrum f\" ur Astronomie, Institut f\" ur Theoretische Astrophysik, Albert-Ueberle-Str. 2, 69120 Heidelberg, Germany.\\
$^{4}$Universit\" at Heidelberg, Interdisziplin\" ares Zentrum f\" ur Wissenschaftliches Rechnen, Im Neuenheimer Feld 205, 69120 Heidelberg, Germany.\\
$^{5}$Observatoire de Gen\`eve, Universit\'e de Gen\`eve, chemin des Maillettes 51, CH-1290 Sauverny, Switzerland.}

\begin{document}

\date{Accepted XXXX December XX. Received XXXX December XX; in original form XXXX October XX}

\pagerange{\pageref{firstpage}--\pageref{lastpage}} \pubyear{2002}

\maketitle

\label{firstpage}

\begin{abstract}

Models aiming to explain the formation of massive black hole seeds, and in particular the direct collapse scenario, face substantial difficulties. 
These are rooted in rather ad hoc and fine-tuned initial conditions, such as the simultaneous requirements of extremely low metallicities and strong radiation backgrounds. 
Here we explore a modification of such scenarios where a massive primordial star cluster is initially produced.
Subsequent stellar collisions give rise to the formation of massive ($10^4$ - $10^5$ $\rm{M}_\odot$) objects. 
Our calculations demonstrate that the interplay between stellar dynamics, gas accretion and protostellar evolution is particularly relevant. 
Gas accretion onto the protostars enhances their radii, resulting in an enhanced collisional cross section. 
We show that the fraction of collisions can increase from 0.1-1\% of the initial population to about 10\% when compared to gas-free models or models of protostellar clusters in the local Universe.   
We conclude that very massive objects can form in spite of initial fragmentation, making the first massive protostellar clusters viable candidate birth places for observed supermassive black holes. 

\end{abstract}


\section{Introduction}\label{sec:intro}

More than 100 supermassive black holes have already been detected at $z>6$ \citep{Gallerani17}, with the number still continuously increasing, 
including 32 quasars recently discovered between $z=5.7$ and $6.8$ via Subaru \citep{Matsuoka17} and 8 $z>6$ quasars via SED model fitting of VISTA, WISE and Dark Energy Survey Year 1 observations \citep{Reed72}. The currently most distant known quasars are at $z=7.085$, i.e. $0.77$ billion years after the Big Bang, with a mass of about $2\times10^9$~M$_\odot$ \citep{Mortlock11}, { and at $z=7.54$ with a mass of about $8\times10^8$~M$_\odot$ \citep{2017arXiv171201860B}.} 

Explaining the existence of these objects provides a significant challenge to our cosmological model, as the accretion at an Eddington rate requires initial seed masses of order $10^4$~M$_\odot$, when realistic spin parameters and accretion disk models are taken into account \citep{Shapiro05}. The only solutions are very massive seeds or extended periods of super-Eddington accretion, potentially also combinations of both
during the formation and early growth of massive black holes. The possible scenarios leading to their formation were already outlined by
\citet{Rees84}, including the direct collapse of massive gas clouds either to a black hole or to a supermassive star, which later collapses to a black hole via general relativistic instabilities, or alternatively through the formation of a dense stellar cluster, which may either collapse into a black hole through relativistic instabilities, or evolve due to stellar mergers.  

From these, the direct collapse model was often considered as the most promising scenario, as it can potentially produce the most massive seeds. Their formation has been proposed through low-angular momentum material in the first pregalactic halos \citep{Koushiappas04}, through efficient angular momentum transport by gravitational instabilities \citep{Begelman06, Lodato06} or through bars-within-bars instabilities \citep{Begelman09}. Numerical simulations probing the formation of massive black holes via direct collapse were indicating that the latter is possible only if cooling is efficiently suppressed, for instance if the formation of molecular hydrogen is photodissociated via a strong radiation background \citep{Bromm03}. Cosmological hydrodynamics simulations following such a collapse over many orders of magnitude indeed found signatures of self-gravitational instabilities in the regime of atomic hydrogen cooling \citep{Wise08}. An update of the pathways leading to massive black hole formation was given by \citet{Regan09}, considering the cosmological conditions that may help to keep the gas metal-free, thereby preventing strong fragmentation events. 

The first systematic fragmentation study in the atomic cooling regime has been pursued by \citet{Latif13a}, showing that fragmentation indeed can be strongly suppressed if cooling is only feasible via atomic hydrogen lines \citep[see also][for the role of the atomic line cooling transitions]{Schleicher10}. The large accretion rates of $0.1$~M$_\odot$~yr$^{-1}$ or more found in the simulations strongly favor protostellar models with cool atmospheres and highly extended envelopes \citep{Hosokawa13, Schleicher13, 2017ApJ...842L...6W, 2017arXiv170509301H}, implying that feedback is rather inefficient. Under such conditions, final black hole masses of $10^5$~M$_\odot$ can be reached based on the results of cosmological simulations \citep{Latif13b, 2016ApJ...830L..34U}.

Such a scenario however represents the most optimistic case. In particular, one needs to investigate the strength of the radiation background to keep the gas atomic, which is typically described through the strength of the radiation background at $13.6$~eV and parametrized via $J_{21}$, with a value of $1$ corresponding to $10^{-21}$~erg~s$^{-1}$~cm$^{-3}$~Hz$^{-1}$~sr$^{-1}$.The first numerical investigations suggested a critical value of $J_{21}\sim100$ to prevent the formation of molecular hydrogen, while updated chemical networks and more realistic models for the radiation background \citep[see e.g.][]{Sugimura14, Agarwal15} have led to much larger critical values of the order $10^5$ when applied in cosmological simulations \citep[][see also \citet{2017MNRAS.468L..82A} for a discussion of the impact of the spectral shape]{Latif14, Latif15}. Under the threshold, massive stars may potentially still form, even though the mass may be reduced by factors of $10-100$ \citep{Latif14b}.

In addition to molecular hydrogen line cooling, fragmentation can also be induced via metals or dust grains \citep{Omukai08}. In the case of metal line cooling, a metallicity of $10^{-3}$~Z$_\odot$ can already increase the cooling and trigger fragmentation within cosmological simulations \citep{Bovino14}, while even lower metallicities of $10^{-5}$~Z$_\odot$ are sufficient when dust cooling is considered \citep{Dopcke13, Schneider06, Bovino16, Latif16}. The need to both have very strong radiation backgrounds, while keeping the gas metal free, leads to a strong needed of fine-tuning, which at best can be satisfied under very rare conditions \citep[e.g.][]{Agarwal17}, while in fact the need for large values of $J_{21}$ provides a problem for the direct collapse scenario \citep{Dijkstra14}.

Surprisingly, the alternative pathway of black hole formation through stellar clusters has been investigated to a lesser degree in the context of early Universe black hole formation. Analytical models by \citet{Devecchi09} and \citet{Devecchi12} predict black hole masses of $100-1000$~M$_\odot$ forming in the first stellar clusters. \citet{Katz15} model the formation of a dense stellar cluster in a cosmological simulation, and show the subsequent formation of a $\sim1000$~M$_\odot$ black hole via $N$-body simulations. Similarly, \citet{Sakurai17} showed the formation of black holes with $400-1900$~M$_\odot$ via a combination of cosmological and $N$-body simulations. So far, these simulations have not considered the enhanced  protostellar radii in the presence of accretion \citep{Hosokawa13, Schleicher13}, and they also neglected the initial presence of gas in the cluster after the formation of the protostars. As we will show in this paper, their ongoing accretion events may however considerably favor collisions and mergers, and thus the formation of a central massive object.

For present-day protostellar clusters, \citet{2011MNRAS.413.1810B} have shown that $0.1-1\%$ of the protostars in the cluster may collide and help to form a particularly massive star within the cluster \citep[see also][for similar results]{Moeckel11, Oh12, Fujii13}. For star clusters at high redshift, we can however expect significantly higher collision rates. One reason is that these clusters are more dense, as the trace amount of metals will trigger cooling and fragmentation only when high densities of order $10^9$~cm$^{-3}$ are reached, thereby leading to the formation of very compact star clusters \citep{2008ApJ...672..757C, 2011Sci...331.1040C, 2011ApJ...737...75G}. Due to the low metallicity and the larger gas temperatures, also the accretion rates will be enhanced, thus favoring collisions through the accretion and mass growth itself, but also due to the protostellar radii, that are enhanced as a result \citep{2012MNRAS.424..457S}. 

We present here the first investigation which explores the formation of massive black hole seeds from a dense stellar cluster, where gas-phase effects like accretion as well as the resulting enhanced protostellar radii are taken into account. We describe our numerical methods in Sec.~\ref{sec:methods} and our experimental setup in Sec.~\ref{sec:exp_setup}. The validation is described in Sec.~\ref{app:B}. Our results are presented in Sec.~\ref{sec:results}, including both our reference model as well as an exploration of the parameter dependence.We summarize the main conclusions in Sec.~\ref{sec:conclusion}.

\section{Numerical Methods}\label{sec:methods}

Modelling the early evolution of a Population III (Pop.~III) protostar cluster is challenging. The main reason is the variety of physical processes that play a role, i.e. gravitational $N$-body dynamics, gravitational coupling between the stars and the gas, stellar growth in mass and size due to gas accretion, and stellar collisions. In this section we describe each of these physical ingredients that go into our simulations, and how we couple them into one numerical model. We use the Astrophysical Multi-purpose Software Environment \citep[\texttt{AMUSE}\footnote{http://www.amusecode.org/},][]{2009NewA...14..369P, PortegiesZwart2013456, AMUSE13}, which was designed for performing such multi-physics simulations as required for our study. Particularly it has the flexibility to introduce new physical ingredients, such as a mass-radius parametrization for accreting Pop.~III protostars, and to couple it to existing $N$-body codes and background potentials \citep[e.g.][]{siblings16, 2017MNRAS.471.3590B}.

\subsection{Initial conditions and dynamics}

Our astrophysical system under investigation consists of Pop.~III protostars embedded in their natal gas cloud. We will assume a simplified initial condition: the protostars and the gas are distributed equally, and they both follow the commonly used Plummer distribution \citep{1911MNRAS..71..460P}. In order to have a well-defined size of the cluster and to ensure that each protostar starts out within the gas cloud, we introduce a cut-off radius after which the density is set to zero. We set this radius to five times the Plummer radius so that the cluster remains stable. The parameters specifying the initial conditions are then: the total gas mass, $M_g$, the cut-off radius, $R_g$, and the number of protostars, $N$. The initial mass of the protostars is set to $m_0 = 0.1\,\rm{M}_\odot$. { In appendix~\ref{app:BB} we confirm that the specific choice of initial mass does not change the results much, as long as we start out with a gas-dominated system}. Complicating factors such as a flattened distribution, cluster rotation and an initial binary fraction are not taken into account in the current study. Our main aim is to explore the complex interplay between dynamics and accretion that will lead to the collisional growth of a massive object.  

To model the star-star gravitational interactions we use the $N$-body code \texttt{ph4} \citep[e.g.][Sec.~3.2]{1996ApJ...467..348M}, which is a fourth-order Hermite algorithm in combination with the time-symmetric integration scheme of \citet{1995ApJ...443L..93H}. We simplify the gravitational dynamics of the gas cloud by an analytical background potential. This potential is coupled to the stars using the \texttt{BRIDGE} method \citep{bridge07}, so that the stars experience both the gravitational force from each other as well as from the gas. Especially at the start of the simulation, when the protostars are still low mass, their orbits will be completely determined by the dominant background potential. 

\subsection{Gas accretion models}\label{sec:accretion_models}

\begin{table}
\centering
\begin{tabular}{llll}
\hline
Model & Gas reservoir & Position dependent & Time dependent \\
      &               & accretion model    & accretion model\\
\hline 			
1 & Infinite & no  & no \\
2 & Infinite & yes & no \\
3 & Finite   & no  & no \\
4 & Finite   & yes & no \\
5 & Finite   & no  & yes \\
6 & Finite   & yes & yes \\
\hline
\end{tabular}
\caption{ Six different gas accretion models for Pop.~III protostars embedded in their natal gas cloud.  }
\label{tab:accretion_models}
\end{table}

The initially low mass Pop.~III protostars will gain mass by accreting from the gas reservoir. Since the accretion rate might vary with cluster environment and cluster evolution, we define six different accretion models based on: (in)finite gas reservoir, position (in)dependent accretion rate, and time (in)dependent accretion rate. An infinite gas reservoir resembles a system that is constantly being fed fresh gas. This prolongs the gas dominated phase and the time scale on which the protostars can accrete gas. 
This is contrary to the finite gas reservoir models, where the gas will eventually run out, and the protostars will stop accreting. For the position-dependent models we set the accretion rate proportional to the local gas density. In this way the protostars in the core accrete at a higher rate than protostars in the halo.  
Such a system naturally produces a range of stellar masses and radii.
In order to compare the position (in)dependent models, we make sure that the cumulative accretion rate is initially equal. A time dependence to the accretion rate is introduced if we set the accretion rate proportional to the gas density, which in turn decreases in time for the models with a finite gas reservoir. The main effect of a decreasing accretion rate is that the stars will migrate to lower mass-radius tracks, and thus have a decreasing collisional cross section. The different combinations of the accretion properties described above define the six different accretion models presented in Tab.~\ref{tab:accretion_models}. An illustration of the time evolution of the accretion rate in the different models is presented in appendix~\ref{app:BB}.   

\subsection{Mass-radius evolution}\label{sec:mrmode}

\begin{figure*}
\centering
\begin{tabular}{cc}
\includegraphics[height=0.56\textwidth,width=0.4\textwidth]{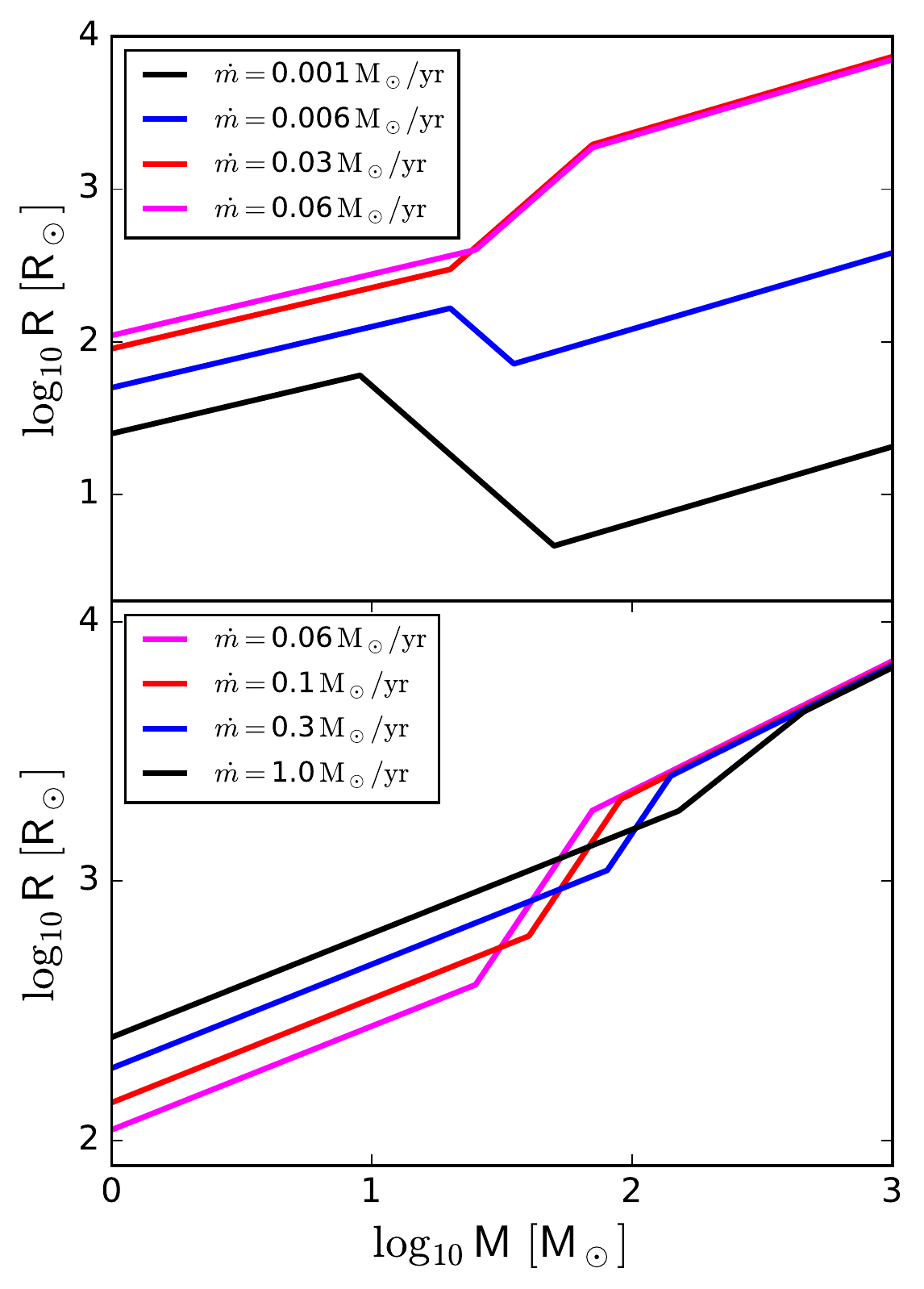} &
\includegraphics[height=0.56\textwidth,width=0.4\textwidth]{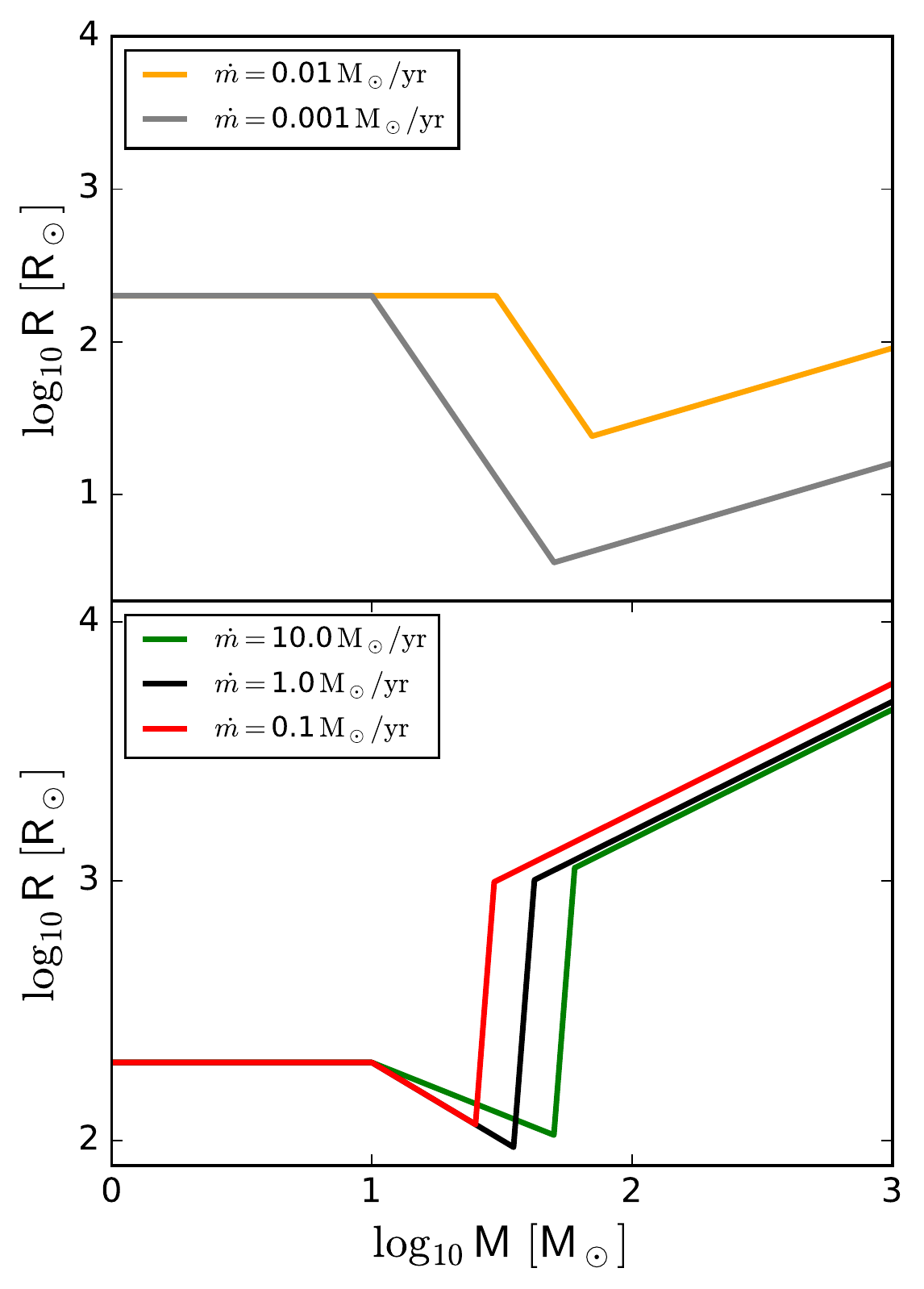} \\
\end{tabular}  
\caption{ Two parametrizations of the mass-radius evolution of accreting Pop.~III protostars based on \citet[][Fig.~5]{Hosokawa2012} (left panel) and \citet[][Fig.~2]{2017arXiv170509301H} (right panel). }
\label{fig:lookup}
\end{figure*}

The mass-radius evolution of accreting Pop.~III protostars is very uncertain, and so we investigate two different sets of models to have a handle on the uncertainties introduced by different protostellar models. 
We base our mass-radius parametrizations on two different studies: \citet[][Fig.~5]{Hosokawa2012} and \citet[][Fig.~2]{2017arXiv170509301H}. Both of these studies performed detailed stellar evolution calculations using  codes by \citet{2003ApJ...589..677O}, \citet{2009ApJ...691..823H}, \citet{2010ApJ...721..478H}, and \texttt{GENEVA} \citep{2008Ap&SS.316...43E} respectively. 

The radius of a protostar is completely determined by its mass, $m$, and accretion rate, $\dot{m}$. At every time step in our simulation we keep track of these two quantities and update the radius of the protostar. For values of the accretion rate in between the parametrized mass-radius tracks, we use interpolation between the two nearest tracks in log-space.  In Fig.~\ref{fig:lookup} we present our approximate parametrization of the mass-radius tracks of accreting Pop.~III protostars, based on the detailed calculations of \citet[][Fig.~5]{Hosokawa2012} and \citet[][Fig.~2]{2017arXiv170509301H}. For the higher accretion rates, the models of \citet{2017arXiv170509301H} produce somewhat smaller radii, but they show the same behaviour at large masses. We use both models to test the sensitivity of the formation of massive objects to the underlying mass-radius parametrization. The analytical form of the parametrization is given appendix~\ref{massradius}.

\subsection{Stellar collisions}

We adopt the commonly used "sticky-sphere" approximation to treat collisions between protostars. 
Whenever the distance between two protostars is less than the sum of their radii, we replace the two protostars by a single object at their center of mass. We assume that during the collision the total mass is conserved, and the new radius is determined by the mass-radius parametrization. In case the accretion rate is very low, i.e. $<10^{-6}\,\rm{M_\odot\,yr^{-1}}$, then the new radius is determined by conserving the density of the primary star.

\section{Experimental Setup}\label{sec:exp_setup}

\begin{table*}
\centering
\begin{tabular}{llllll|llllllllll}
\hline
$M_{\rm{g}}$          & $R_{\rm{g}}$     & $N$ & $\dot{m}$               & $M_{\rm{m,1}}$ & $M_{\rm{m,2}}$ & $M_{\rm{m,3}}$ & $M_{\rm{m,4}}$ & $M_{\rm{m,5}}$ & $M_{\rm{m,6}}$ & $N_{\rm{c,1}}$ & $N_{\rm{c,2}}$ & $N_{\rm{c,3}}$ & $N_{\rm{c,4}}$ & $N_{\rm{c,5}}$ & $N_{\rm{c,6}}$ \\
\hline 			
$\rm{M_\odot}$ & $\rm{pc}$ &     & $\rm{M_\odot\,yr^{-1}}$ & $\rm{M_\odot}$ &  $\rm{M_\odot}$ & $\rm{M_\odot}$ & $\rm{M_\odot}$ & $\rm{M_\odot}$ & $\rm{M_\odot}$ & & & & & & \\
\hline 			
\hline
$10^{5}$ & 0.1 & 256 & 0.03 & 128698 & 122888 & 73463 & 88127 & 13167 & 81490 & 238 & 223 & 196 & 197 & 52 & 186 \\
\hline
$10^{4}$ & 0.1 & 256 & 0.03 & 52567 & 60425 & 4069 & 7997 & 627 & 986 & 242 & 231 & 110 & 124 & 15 & 13 \\
$3\times10^{4}$ & 0.1 & 256 & 0.03 & 68380 & 91207 & 18737 & 26004 & 2240 & 14670 & 239 & 234 & 162 & 182 & 19 & 81 \\
$3\times10^{5}$ & 0.1 & 256 & 0.03 & 165820 & 133309 & 138241 & 179454 & 111502 & 131137 & 231 & 196 & 216 & 198 & 166 & 205 \\
$10^{6}$ & 0.1 & 256 & 0.03        & 179746 & 172681 & 178408 & 172964 & 158720 & 142053 & 230 & 191 & 219 & 191 & 209 & 187  \\
\hline
$10^{5}$ & 0.01 & 256 & 0.03 & 10707 & 9449 & 11734 & 9343 & 12557 & 12406 & 239 & 203 & 239 & 205 & 239 & 210 \\
$10^{5}$ & 0.03 & 256 & 0.03 & 48319 & 43226 & 32766 & 33458 & 36281 & 31088 & 239 & 211 & 226 & 204 & 209 & 191 \\
$10^{5}$ & 0.3 & 256 & 0.03  & 253063 & 362374 & 53793 & 79250 & 5796 & 47880 & 230 & 230 & 140 & 150 & 16 & 81 \\
$10^{5}$ & 1.0 & 256 & 0.03  & 701633 & 1020974 & 38607 & 45433 & 2345 & 4092 & 223 & 207 & 100 & 59 & 7 & 5 \\
\hline
$10^{5}$ & 0.1 & 64 & 0.03   & 48671 & 48945 & 32849 & 56258 & 38327 & 44628 & 53 & 48 & 40 & 44 & 42 & 40 \\
$10^{5}$ & 0.1 & 128 & 0.03  & 87493 & 85516 & 77464 & 73564 & 16884 & 42905 & 114 & 88 & 105 & 81 & 43 & 69 \\
$10^{5}$ & 0.1 & 512 & 0.03  & 155664 & 181666 & 68136 & 86562 & 7507 & 70803 & 500 & 439 & 371 & 362 & 52 & 319 \\
$10^{5}$ & 0.1 & 1024 & 0.03 & 226161 & 275657 & 74954 & 91700 & 6101 & 57747 & 1004 & 985 & 792 & 752 & 68 & 455  \\
\hline
$10^{5}$ & 0.1 & 256 & 0.001 & 23649 & 36107 & 11059 & 28385 & 10773 & 19905 & 62 & 114 & 31 & 91 & 35 & 68 \\
$10^{5}$ & 0.1 & 256 & 0.003 & 58318 & 65221 & 24661 & 47494 & 11610 & 45009 & 104 & 188 & 68 & 149 & 43 & 133 \\
$10^{5}$ & 0.1 & 256 & 0.01  & 97024 & 94974 & 35739 & 79289 & 8204 & 54539 & 174 & 210 & 103 & 191 & 26 & 171 \\
$10^{5}$ & 0.1 & 256 & 0.1   & 195809 & 289809 & 78644 & 88052 & 20661 & 79274 & 245 & 245 & 206 & 192 & 77 & 170 \\
$10^{5}$ & 0.1 & 256 & 0.3   & 244229 & 309201 & 53819 & 89040 & 13292 & 73790 & 245 & 250 & 149 & 174 & 51 & 136 \\
\hline
\end{tabular}
\caption{ Overview of the simulations. The input parameters are: gas cloud mass $M_{\rm{g}}$, gas cloud radius $R_{\rm{g}}$, number of protostars $N$ and the average accretion rate $\dot{m}$. Statistics describing the output are: final mass of the most massive object $M_{\rm{m}}$ and total number of collisions $N_{\rm{c}}$. The number in the subscript denotes the accretion model (see Tab.~\ref{tab:accretion_models}).  }
\label{tab:simulation_overview}
\end{table*}

\begin{table*}
\centering
\begin{tabular}{llllll|llllllllll}
\hline
$M_{\rm{g}}$          & $R_{\rm{g}}$     & $N$ & $\dot{m}$               & $R_{\rm{m,1}}$ & $R_{\rm{m,2}}$ & $R_{\rm{m,3}}$ & $R_{\rm{m,4}}$ & $R_{\rm{m,5}}$ & $R_{\rm{m,6}}$ & $R_{\rm{av,1}}$ & $R_{\rm{av,2}}$ & $R_{\rm{av,3}}$ & $R_{\rm{av,4}}$ & $R_{\rm{av,5}}$ & $R_{\rm{av,6}}$ \\
\hline 			
$\rm{M_\odot}$ & $\rm{pc}$ &     & $\rm{M_\odot\,yr^{-1}}$ & $\rm{T_{cr,0}^{-1}}$ & $\rm{T_{cr,0}^{-1}}$ & $\rm{T_{cr,0}^{-1}}$ & $\rm{T_{cr,0}^{-1}}$ & $\rm{T_{cr,0}^{-1}}$ & $\rm{T_{cr,0}^{-1}}$ & $\rm{T_{cr,0}^{-1}}$ & $\rm{T_{cr,0}^{-1}}$ & $\rm{T_{cr,0}^{-1}}$ & $\rm{T_{cr,0}^{-1}}$ & $\rm{T_{cr,0}^{-1}}$ & $\rm{T_{cr,0}^{-1}}$ \\
\hline 			
\hline
$10^{5}$ & 0.1 & 256 & 0.03        & 21 & 22 & 14 & 19 & 4 & 14 & 1.60 & 3.37 & 0.32 & 1.29 & 0.36 & 0.30 \\
\hline
$10^{4}$ & 0.1 & 256 & 0.03        & 145 & 125 & 8 & 31 & 1 & 8 & 14.35 & 20.17 & 0.20 & 0.08 & 0.02 & 0.43 \\
$3 \times 10^{4}$ & 0.1 & 256 & 0.03 & 55 & 69 & 16 & 62 & 2 & 43 & 9.66 & 10.52 & 0.15 & 0.14 & 0.05 & 4.12 \\
$3 \times 10^{5}$ & 0.1 & 256 & 0.03 & 13 & 9 & 8 & 17 & 6 & 11 & 0.87 & 1.35 & 0.78 & 0.32 & 0.42 & 0.50 \\
$10^{6}$ & 0.1 & 256 & 0.03        & 7 & 5 & 7 & 7 & 6 & 6 & 0.39 & 0.26 & 0.38 & 0.26 & 0.35 & 0.45 \\
\hline
$10^{5}$ & 0.01 & 256 & 0.03 & 11 & 15 & 14 & 13 & 12 & 9 & 1.47 & 0.69 & 0.65 & 0.78 & 0.55 & 0.28 \\
$10^{5}$ & 0.03 & 256 & 0.03 & 15 & 13 & 21 & 17 & 13 & 12 & 0.67 & 0.48 & 1.58 & 1.01 & 0.30 & 0.71 \\
$10^{5}$ & 0.3 & 256 & 0.03  & 47 & 55 & 9 & 49 & 2 & 30 & 5.12 & 4.11 & 0.16 & 0.17 & 0.04 & 4.22 \\
$10^{5}$ & 1.0 & 256 & 0.03  & 83 & 71 & 7 & 23 & 2 & 3 & 7.16 & 5.58 & 0.30 & 0.21 & 0.10 & 0.79 \\
\hline
$10^{5}$ & 0.1 & 64 & 0.03   & 5 & 5 & 6 & 9 & 3 & 3 & 0.45 & 0.53 & 0.48 & 0.38 & 0.26 & 0.25 \\
$10^{5}$ & 0.1 & 128 & 0.03  & 10 & 9 & 7 & 13 & 5 & 7 & 0.77 & 0.38 & 0.90 & 0.64 & 0.52 & 0.66 \\
$10^{5}$ & 0.1 & 512 & 0.03  & 81 & 57 & 48 & 59 & 2 & 54 & 5.66 & 3.86 & 2.36 & 2.91 & 0.28 & 2.18 \\
$10^{5}$ & 0.1 & 1024 & 0.03 & 190 & 185 & 52 & 232 & 3 & 123 & 14.71 & 7.53 & 2.06 & 3.10 & 0.42 & 4.86 \\
\hline
$10^{5}$ & 0.1 & 256 & 0.001 & 2 & 4 & 2 & 3 & 1 & 4 & 0.07 & 0.13 & 0.04 & 0.11 & 0.01 & 0.11 \\
$10^{5}$ & 0.1 & 256 & 0.003 & 3 & 7 & 4 & 5 & 3 & 6 & 0.13 & 0.33 & 0.01 & 0.43 & 0.07 & 0.14 \\
$10^{5}$ & 0.1 & 256 & 0.01  & 7 & 12 & 7 & 12 & 3 & 8 & 0.57 & 0.68 & 0.08 & 0.57 & 0.03 & 0.59 \\
$10^{5}$ & 0.1 & 256 & 0.1   & 42 & 62 & 28 & 44 & 12 & 22 & 6.12 & 1.93 & 0.58 & 1.78 & 0.66 & 1.09 \\
$10^{5}$ & 0.1 & 256 & 0.3   & 80 & 114 & 23 & 72 & 10 & 47 & 14.11 & 17.55 & 1.93 & 0.92 & 0.77 & 7.44 \\
\hline
\end{tabular}
\caption{ Overview of the simulations. Continuing Tab.~\ref{tab:simulation_overview} presenting the maximum collision rate $R_{\rm{m}}$, and the average collision rate $R_{\rm{av}}$.  }
\label{tab:simulation_overview2}
\end{table*}

The input parameters specifying a simulation and their range in values are: 

\begin{itemize}
\item Gas cloud mass, $\rm{M_g = 10^{4}, 3 \times 10^{4}, 10^{5}, 3 \times 10^{5}, 10^{6}\,M_\odot}$
\item Gas cloud radius, $\rm{R_g = 0.01, 0.03, 0.1, 0.3, 1.0\,pc}$
\item Number of protostars, $\rm{N = 64, 128, 256, 512, 1024}$
\item Average accretion rate, $\dot{m}$ = 0.001, 0.003, 0.01, 0.03, 
\item[] 0.1, 0.3\,$\rm{M}_\odot\,\rm{yr}^{-1}$.
\end{itemize}

\noindent We also vary the accretion model (see Sec.~\ref{sec:accretion_models}) and the mass-radius parametrization (see Sec.~\ref{sec:mrmode}). 
{ We note that the adopted protostellar accretion rates are high compared to present-day star formation. However, they are realistic for the primordial case. Similar to current star formation, the protostellar accretion process in Pop. III clusters is not regulated via Bondi-Hoyle-Littleton accretion, but rather it has been shown that gravitational instabilities in the gas are driving fragmentation as well as the accretion process onto the fragments. This leads to the typical range of accretion rates under Pop. III conditions that we adopted, see e.g. \citet{2011Sci...331.1040C}, \citet{2011ApJ...737...75G}, \citet{2012MNRAS.424..457S}, \citet{Latif13b}, \citet{2015MNRAS.449...77L}, \citet{2015A&A...578A.118L}, \citet{2015MNRAS.448..568H}, \citet{Regan16}, \citet{2016ApJ...824..119H} and \citet{Latif16}}.
We define a standard set of parameters as: $M_{\rm{g}} = 10^{5}\,\rm{M_\odot}$, $R_{\rm{g}} = 0.1\,\rm{pc}$, $N = 256$, $\dot{m} = 0.03\,\rm{M_\odot\,yr^{-1}}$, and mass-radius parametrization based on \citet{Hosokawa2012}. This choice of parameter values reflects that we are particularly interested in very massive Pop.~III protostar clusters and the formation of very massive objects. The initial crossing time of our systems is given by 

\begin{equation}
T_{\rm{cross,0}} = 853\,\left( \frac{10^5\,\rm{M}_\odot}{M_{\rm{g}}} \right)^{\frac{1}{2}} \left(\frac{R_{\rm{g}}}{0.1\,\rm{ pc}}\right)^{\frac{3}{2}}\,\rm{yr},
\label{eq:tcross}
\end{equation}

\noindent where we used $T_{\rm{cross,0}} = 2 R_{\rm{v}} / \sigma$, with $R_{\rm{v}}$ the virial radius given by $R_{\rm{v}} = 16 R_{\rm{g}} / 15 \pi$ (using the definition of our gas cloud truncation radius), $\sigma$ the velocity dispersion estimated by $\sigma = \sqrt{G M_{\rm{g}} / 2 R_{\rm{v}}}$, and $G$ the gravitational constant. 

We consider a simulation finished if most of the collisions have occurred. We determine this by keeping track of the average collision rate,

\begin{equation}
R_{\rm{av}}\left( t \right) = \frac{N_{\rm{col}}\left( t \right)}{t_{\rm{last\,collision}}},
\end{equation}

\noindent and an upper limit of the current collision rate, 

\begin{equation}
R\left( t \right) = \frac{1}{t-t_{\rm{last\,collision}}}.
\end{equation} 

\noindent If the ratio of $R / R_{\rm{av}} < 0.015$ we stop the simulation. This criterion was chosen to make sure that the majority of collisions have occurred, while also limiting the duration of the simulation. We also stop the simulation if no collision has occurred in the last million years. 

For each simulation we store regular snapshots of the full phase space information, allowing us to retrace the mass-radius evolution and calculate collision rates. In Tab.~{\ref{tab:simulation_overview} and \ref{tab:simulation_overview2} we provide an overview of our simulations and their input parameters, together with several statistics describing the outcome of the simulations, such as the maximum mass and maximum collision rate. An estimate of the measurement uncertainties can be found in Tab.~\ref{tab:mrmode_comparison}. 

In the next section we provide several validation tests of our experimental setup. We show that the initial condition is a stable Plummer sphere distribution (Fig.~\ref{fig:validation_dynamics}), and we validate both the correct implementation of the six different accretion models, and the mass-radius parametrization based on \citet{Hosokawa2012}.

\section{Validation of the numerical method}\label{app:B}

We describe the numerical methods used in this study in Sec.~\ref{sec:methods}. Here we present validation experiments of our simulation setup and numerical implementation.   

We have performed a validation experiment of the initial conditions and dynamics by choosing the following parameters: $M_g = 10^5\,\rm{M_\odot}$, $R_g = 0.1\,\rm{pc}$ and $N = 256$. We evolved the system for a time $T=10^5\,\rm{yr}$ and note that accretion and collisions are not included yet. In Fig.~\ref{fig:validation_dynamics} we confirm that the 10, 50 and 90\% Lagrangian radii follow those of a Plummer sphere, with a slight discrepancy in the 90\% radius due to the truncation that we have introduced. We also note that all the protostars remain within the gas cloud, i.e. there are no escapers due to the initial condition, and the velocity dispersion of the protostars is close to the analytical value of $\rm{\sim 80\,km/s}$. We have constructed a stable star cluster which will only alter its configuration due to gas accretion and stellar collisions.

We have performed a numerical validation experiment by evolving the same initial condition as above, but this time only with gas accretion, i.e. dynamics is turned off. We set the initial accretion rate $\dot{m} = 0.03\,\rm{M_\odot\,yr^{-1}}$, i.e. here and in the rest of this study $\dot{m}$ refers to the average accretion rate per star over the whole cluster. In Fig.~\ref{fig:validation_accretion} we present the time evolution of the total star and gas mass (top row) and the time evolution of the average accretion rate (bottom row). We note that the position dependent and independent accretion models are consistent on average (their curves overlie). Together all these different models cover a variety of physical regimes. 

\begin{figure*}
\centering
\begin{tabular}{ccc}
\includegraphics[height=0.26\textwidth,width=0.3\textwidth]{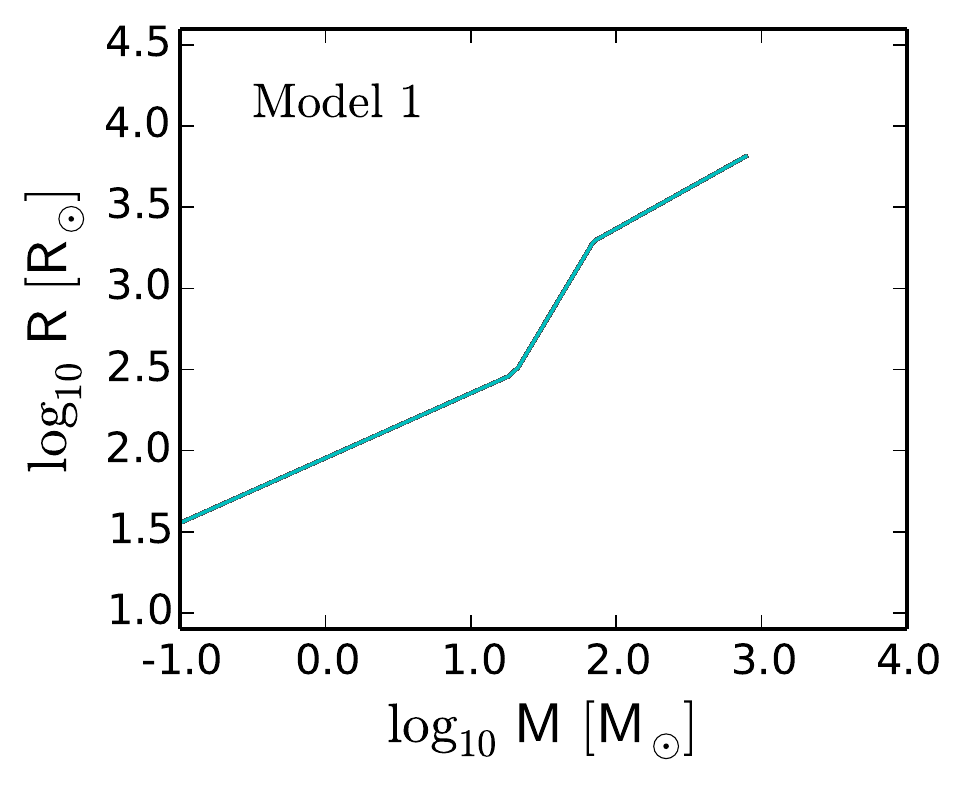} &
\includegraphics[height=0.26\textwidth,width=0.3\textwidth]{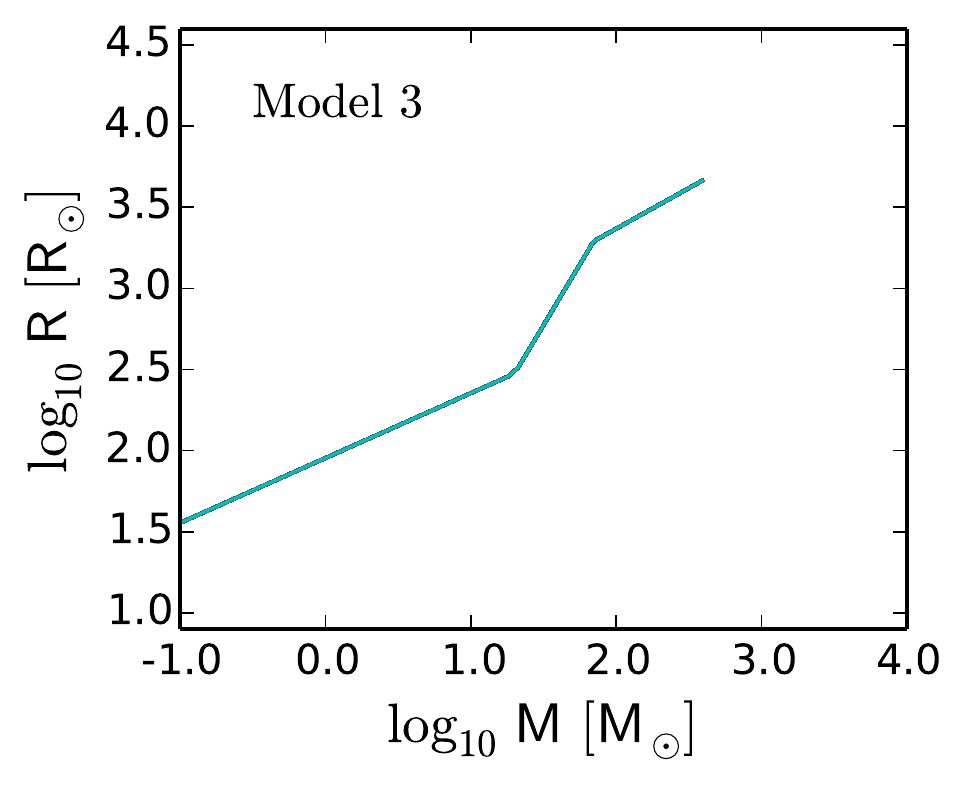} &
\includegraphics[height=0.26\textwidth,width=0.3\textwidth]{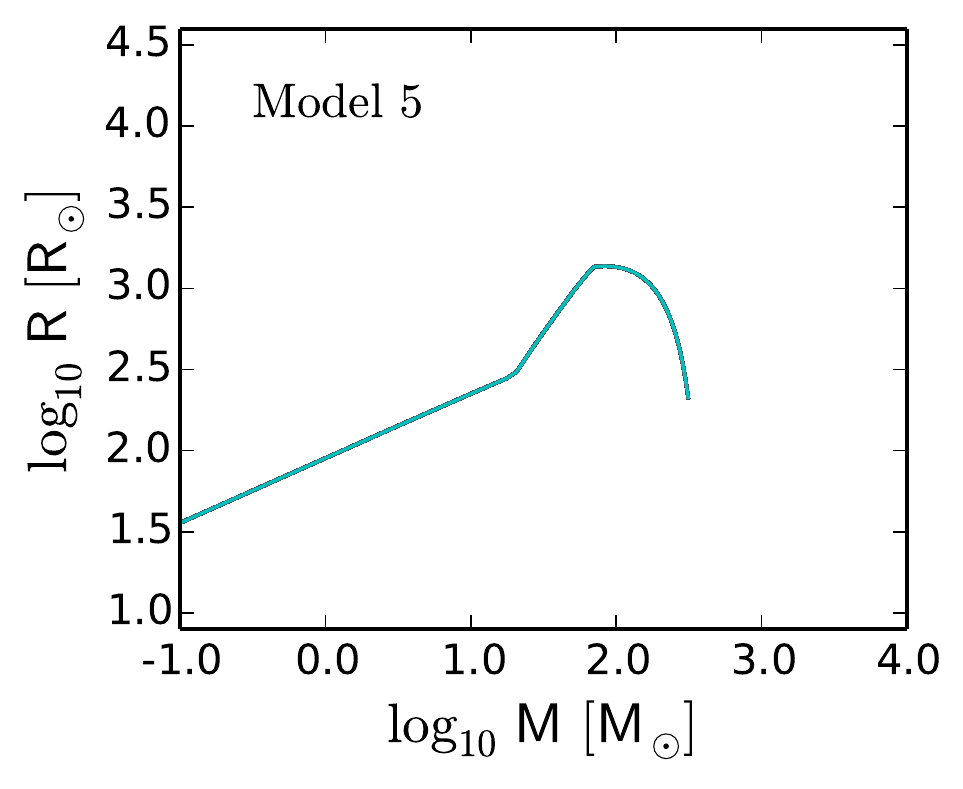} \\

\includegraphics[height=0.26\textwidth,width=0.3\textwidth]{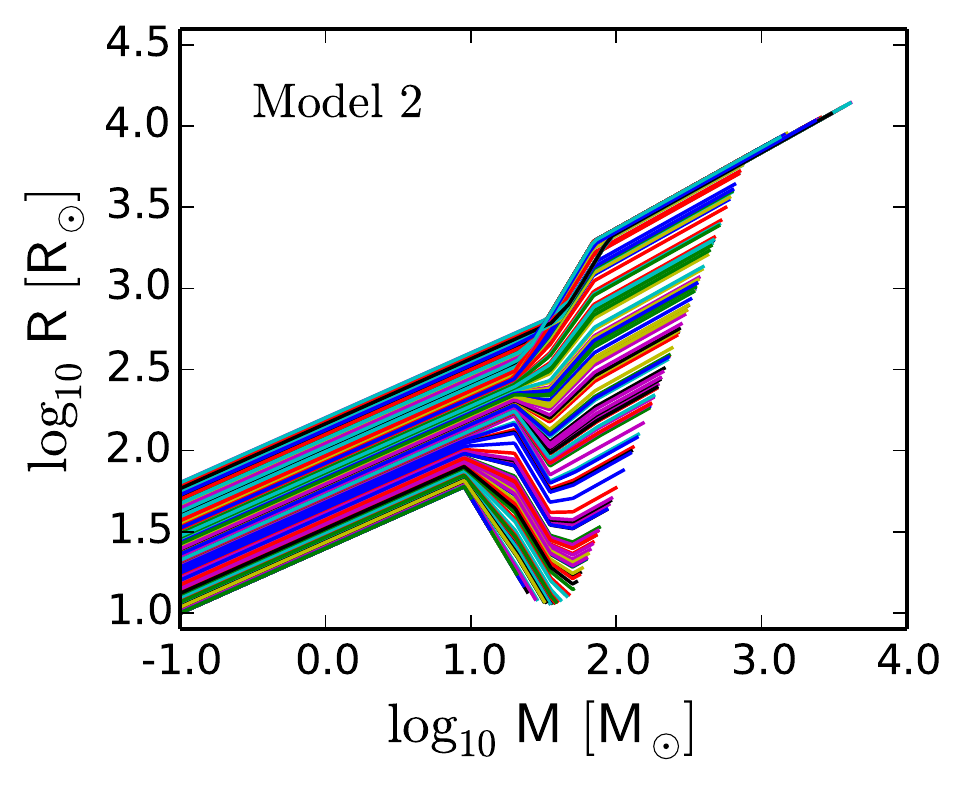} &
\includegraphics[height=0.26\textwidth,width=0.3\textwidth]{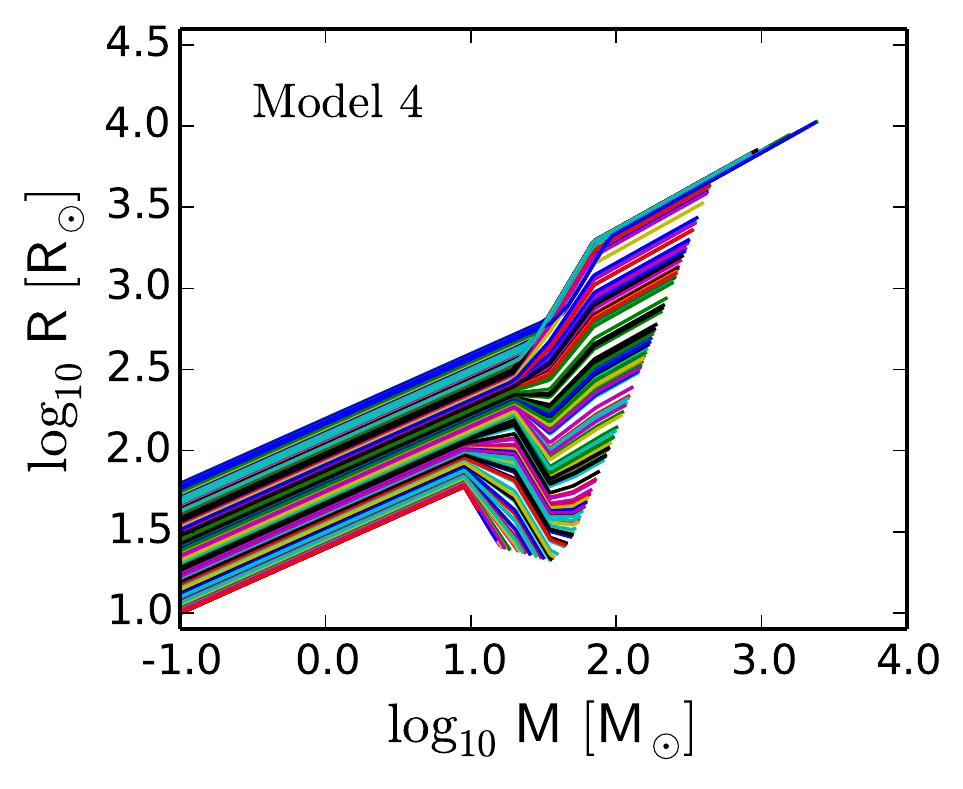} &
\includegraphics[height=0.26\textwidth,width=0.3\textwidth]{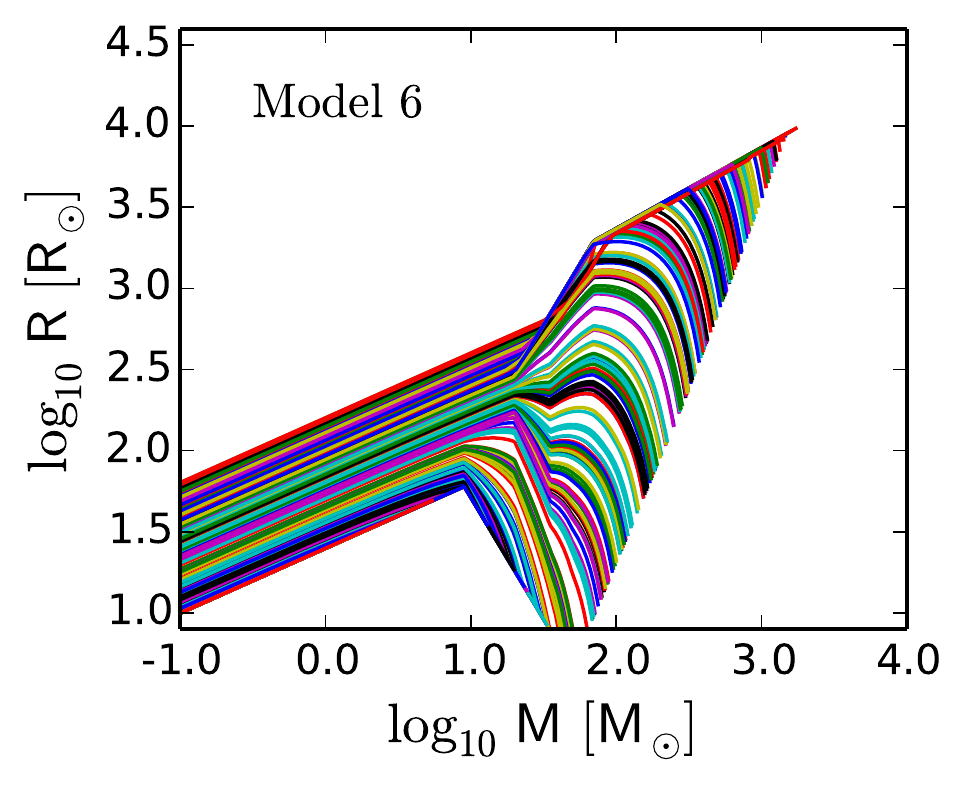} \\
\end{tabular}  
\caption{ Validation of the mass-radius parametrization implementation. We consider a Plummer distribution of protostars, which are accreting gas, resulting in evolving masses and radii. The protostars do not move in this validation experiment, which explains the regularity in the distribution of mass-radius tracks. We confirm that for the position independent accretion models (top row) the protostars follow the mass-radius parametrization based on Hosokawa et al. 2012. For the time dependent accretion model (right panel) the accretion rate decreases in time, so that the protostars migrate to mass-radius tracks of lower accretion rate, causing the protostars to eventually shrink again. For the position dependent accretion models (bottom row) we confirm a correct interpolation of the mass-radius tracks in log-space, and the production of a range of masses and radii. }
\label{fig:validation_mrmode}
\end{figure*}

In the same validation experiment, we also kept track of the mass-radius evolution of the protostars, which we present in Fig.~\ref{fig:validation_mrmode} (for the model of Hosokawa et al. 2012). There we clearly observe the difference between the position independent (top row) and dependent (bottom row) models. The latter produces a spectrum of masses and radii. We note that in the time-dependent models (5 and 6), the accretion rate decreases in time, and as a result the protostars will migrate to lower mass-radius tracks. After an initial phase of growth these protostars will eventually shrink, which is expected to decrease the collision rate. 

\section{Results}\label{sec:results}

We start by analyzing the simulations with the standard set of parameters defined in Sec.~\ref{sec:exp_setup}. We show a proof of concept that many accretion-induced collisions can lead to the formation of a single massive object. Next, we determine the sensitivity of this result to variations of the input parameters. 

\subsection{Simulations with the standard set of parameters}\label{results:standard}

\begin{figure*}
\centering
\begin{tabular}{ccc}
\includegraphics[height=0.26\textwidth,width=0.3\textwidth]{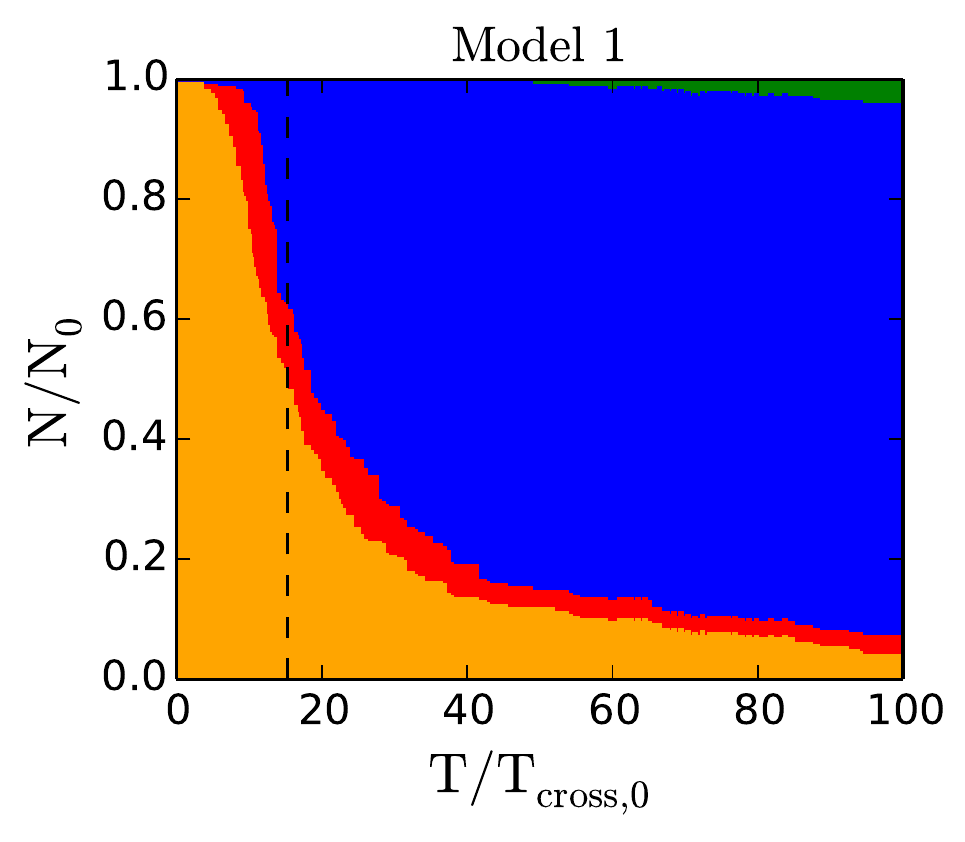} &
\includegraphics[height=0.26\textwidth,width=0.3\textwidth]{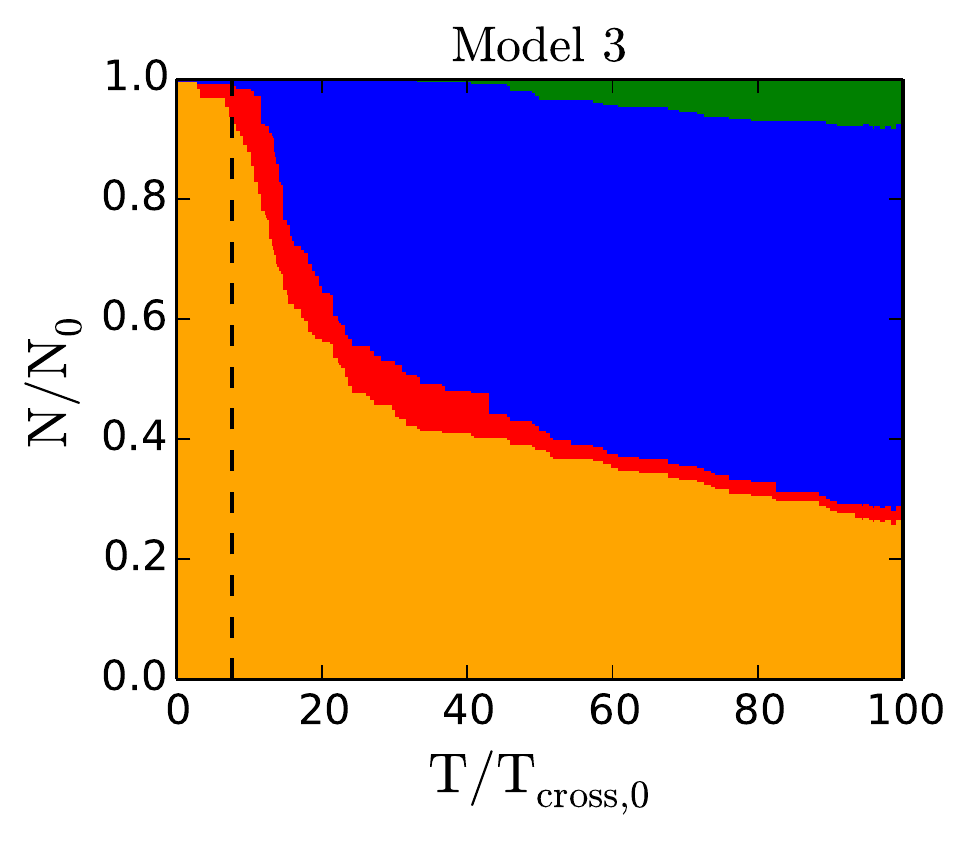} &
\includegraphics[height=0.26\textwidth,width=0.3\textwidth]{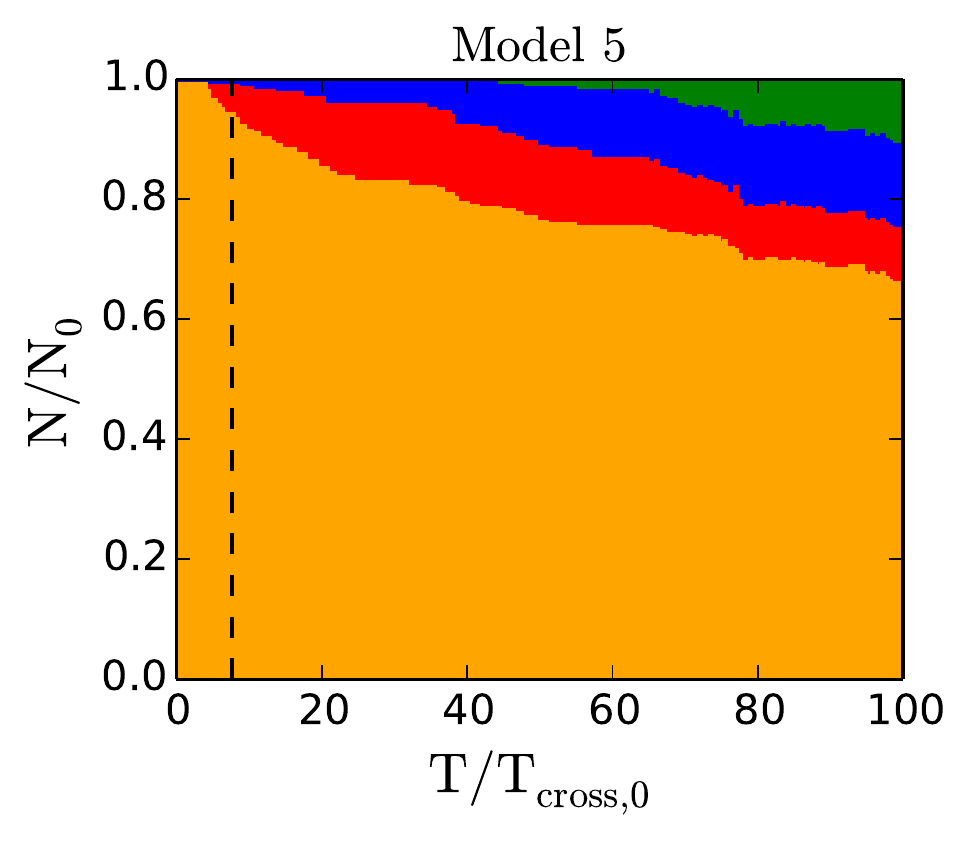} \\

\includegraphics[height=0.26\textwidth,width=0.3\textwidth]{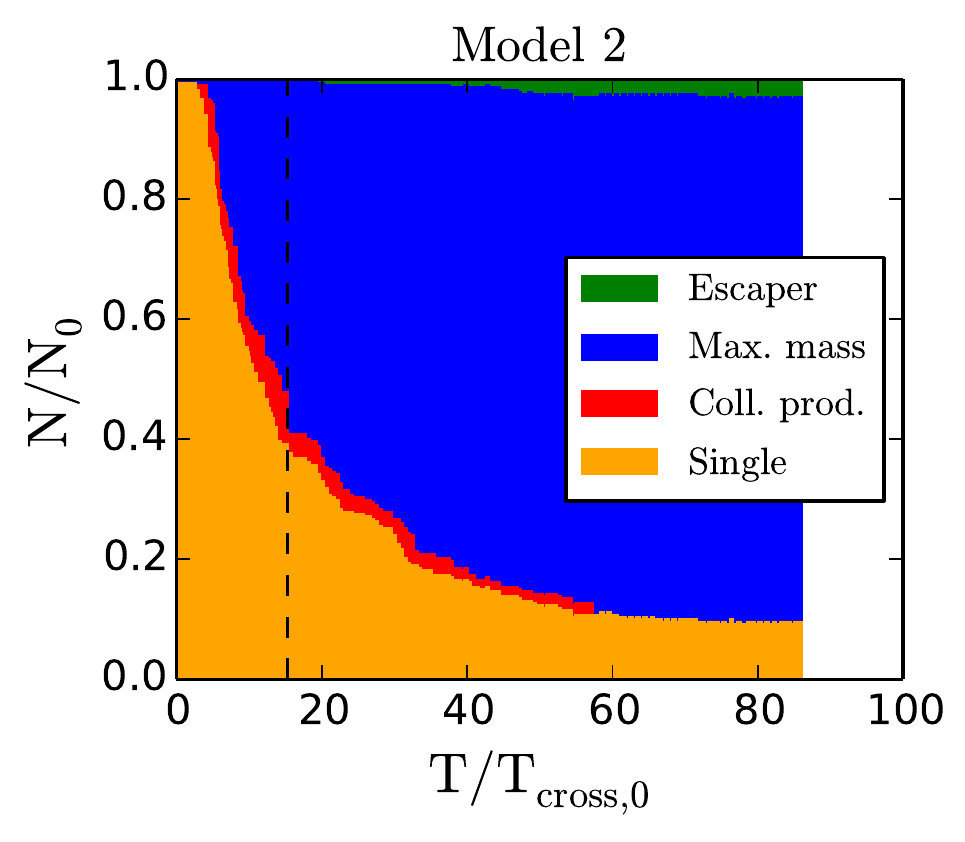} &
\includegraphics[height=0.26\textwidth,width=0.3\textwidth]{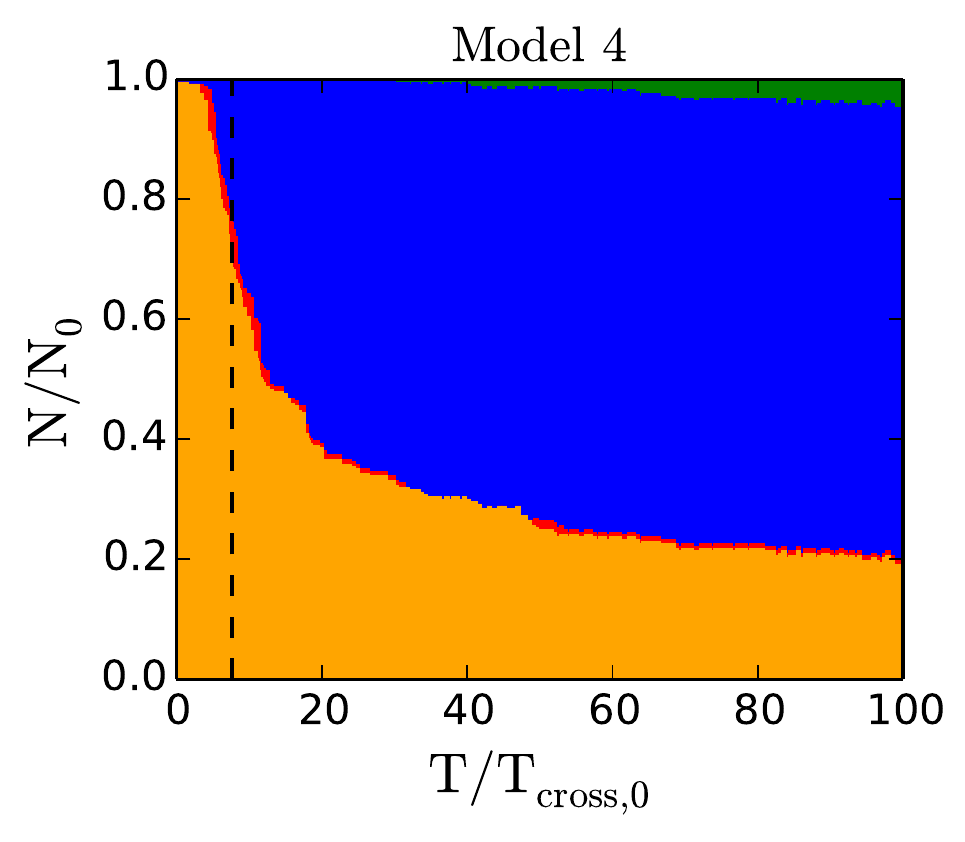} &
\includegraphics[height=0.26\textwidth,width=0.3\textwidth]{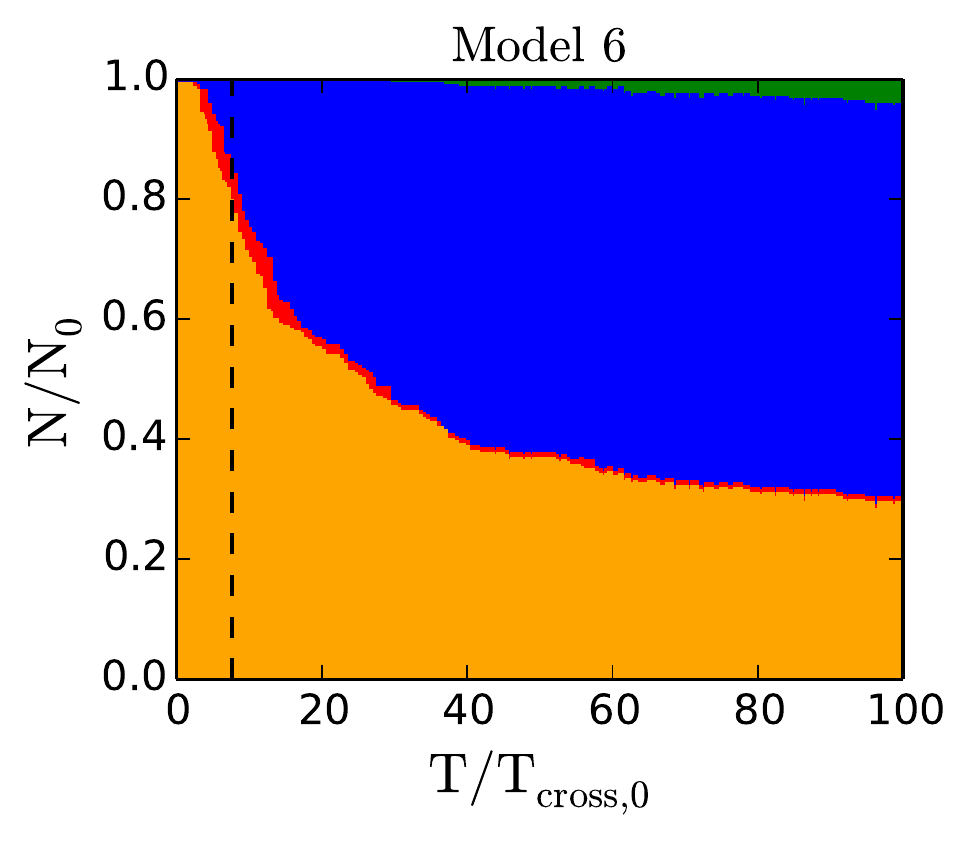} \\
\end{tabular}  
\caption{ Time evolution of the fraction of stars belonging to four categories: 1) escapers (green), 2) stars that have collided with and are part of the most massive star in the system (blue, ``Max. mass''), 3) stars that are part of other collision products (red, ``Coll. prod.''), and 4) single stars (orange). 
We indicate the moment of transition to a stellar mass dominated system (see Sec.~\ref{sec:accretion_models}) with a vertical dashed line.
Initially all the stars are single stars in the system. As the stars accrete gas, they become larger and will dynamically interact with the other stars. We generally observe a decrease in the fraction of single stars, and an increase in collision products, and at late times a small fraction of escapers. }
\label{fig:relative_fractions}
\end{figure*}

\begin{figure*}
\centering
\begin{tabular}{ccc}
\includegraphics[height=0.96\textwidth,width=0.32\textwidth]{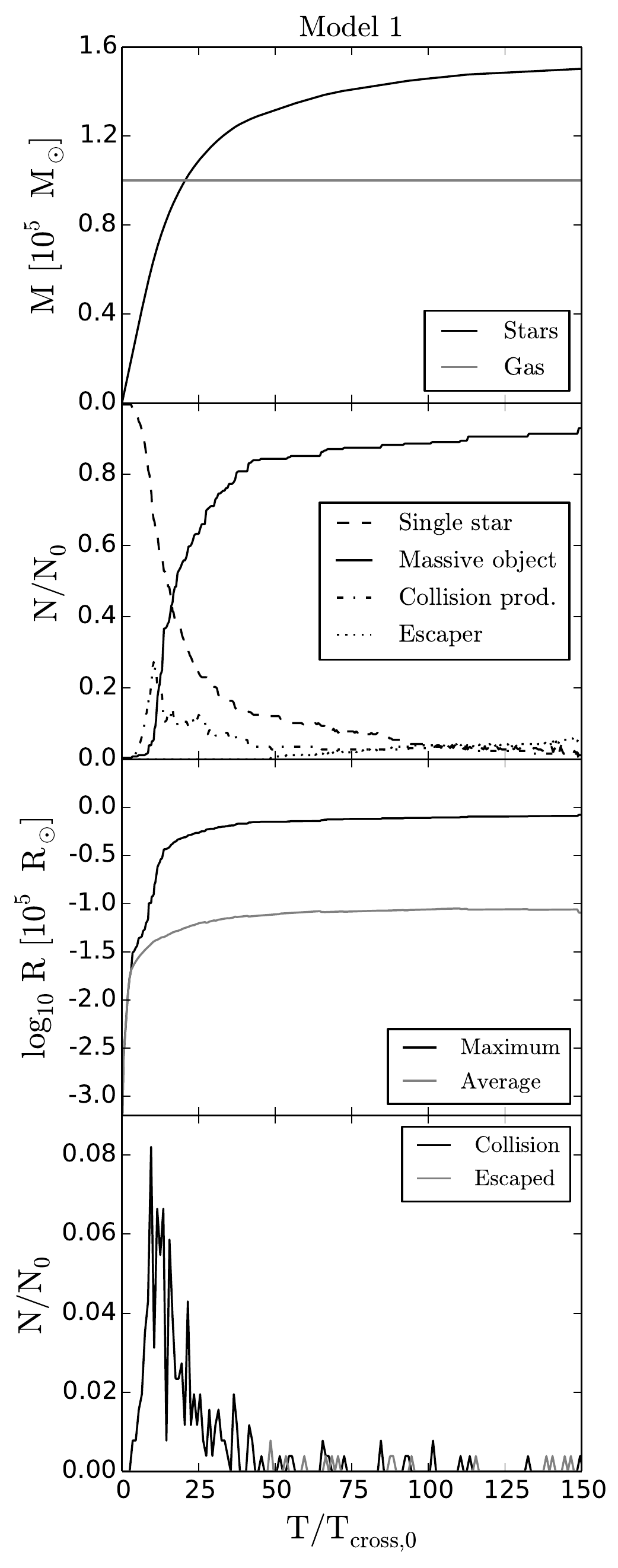} &
\includegraphics[height=0.96\textwidth,width=0.32\textwidth]{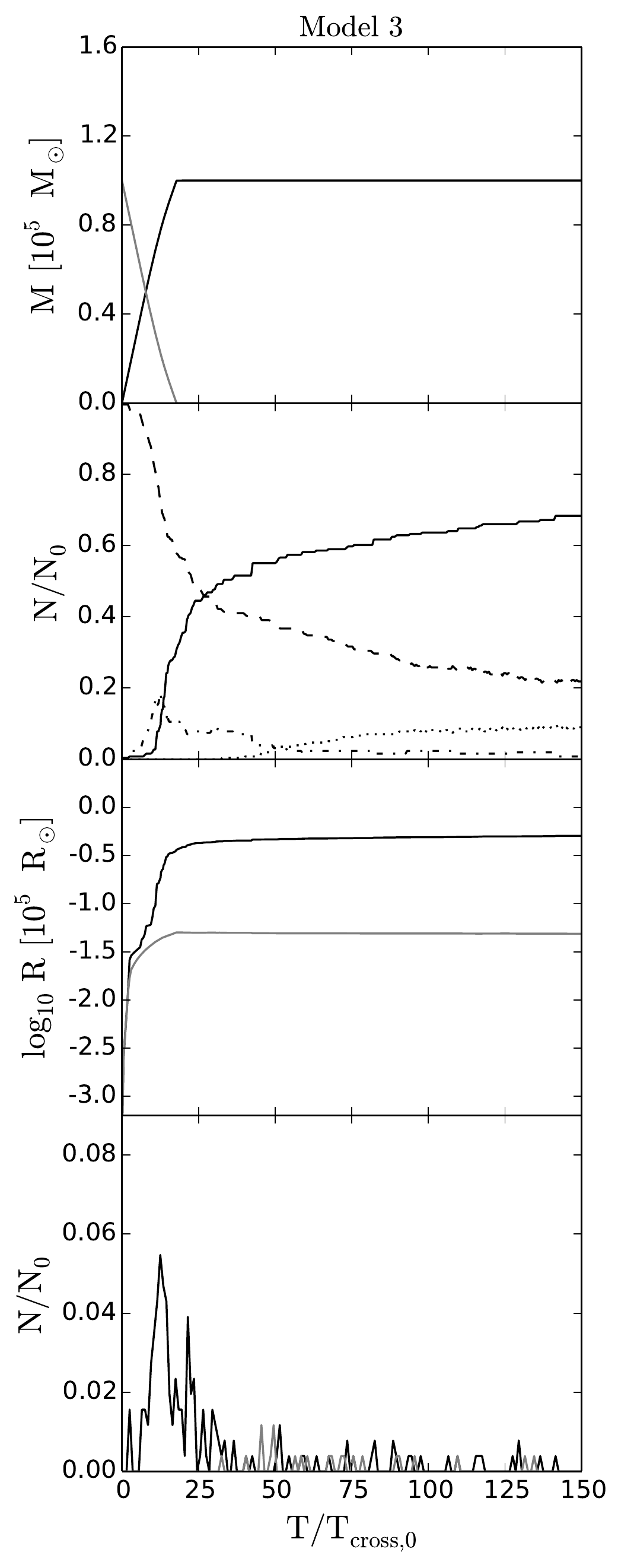} &
\includegraphics[height=0.96\textwidth,width=0.32\textwidth]{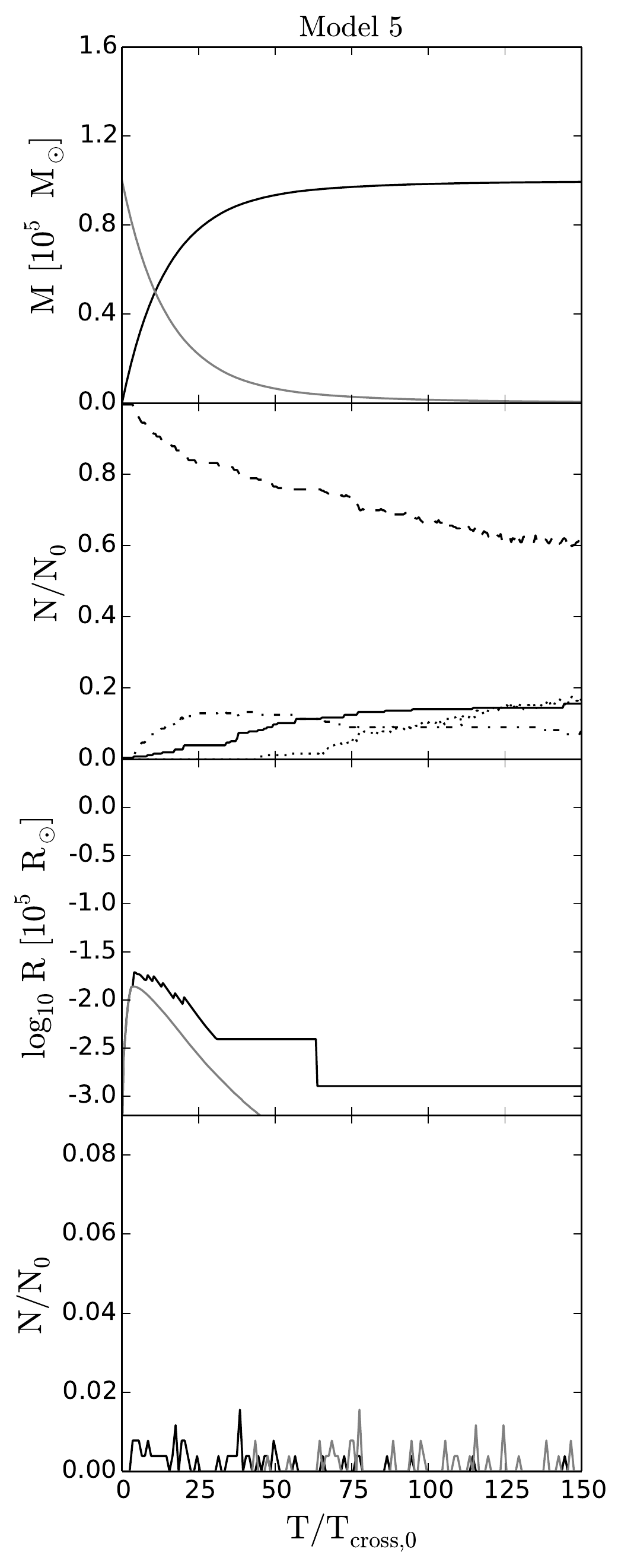} \\
\end{tabular}  
\caption{ Correlating the time evolution of the collision and escape rate (bottom panel) with: total star and gas mass (top panel), fraction of stars belonging to the same four categories as in Fig.~\ref{fig:relative_fractions} (second panel),  and maximum stellar radius and average stellar radius of the remaining stars (third panel). We show these results for the different accretion models (per column, see also next figure). }
\label{fig:time_evolutions}
\end{figure*}

\begin{figure*}
\centering
\begin{tabular}{ccc}
\includegraphics[height=0.96\textwidth,width=0.32\textwidth]{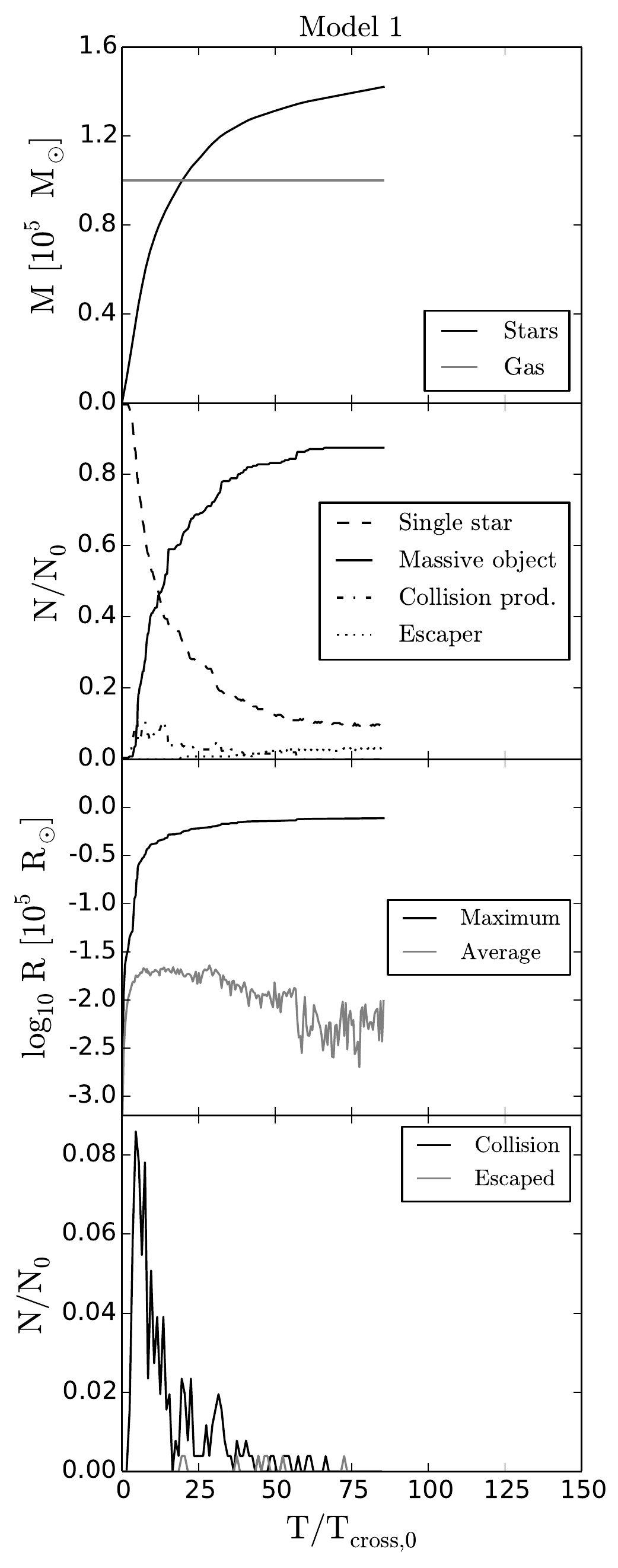} &
\includegraphics[height=0.96\textwidth,width=0.32\textwidth]{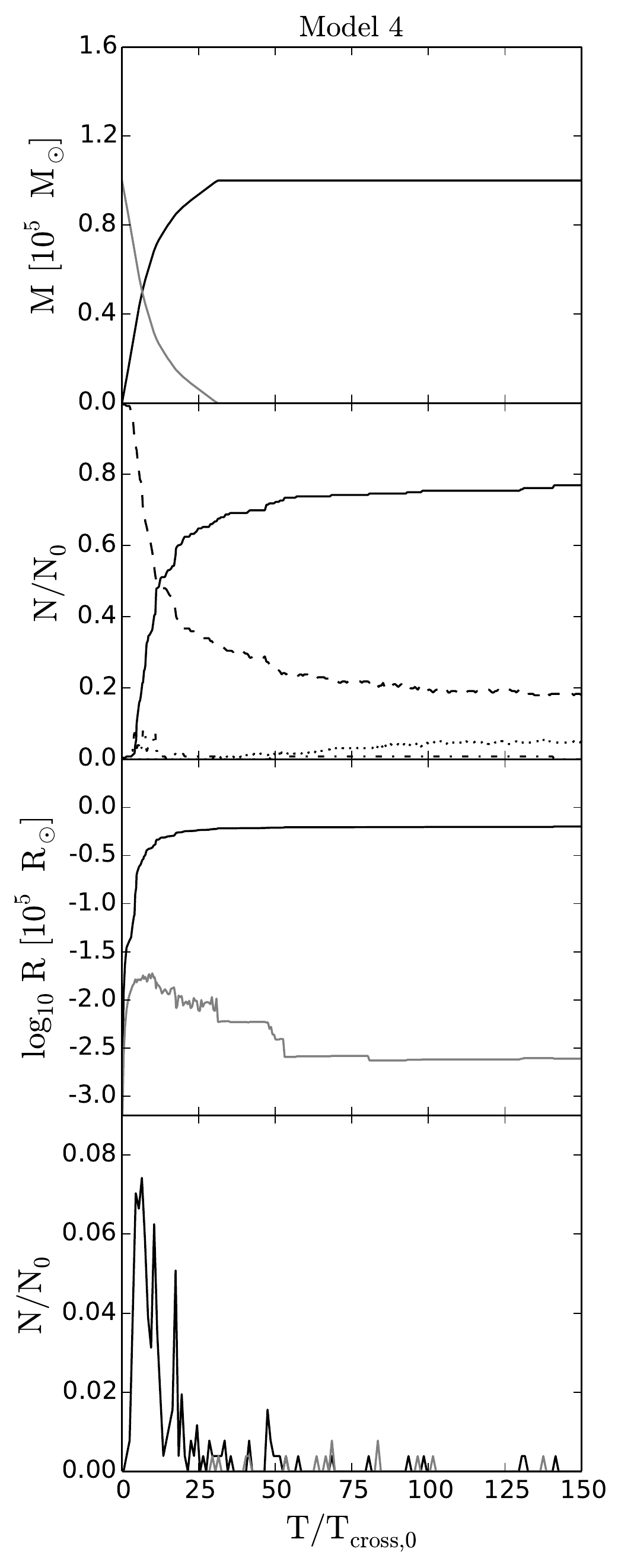} &
\includegraphics[height=0.96\textwidth,width=0.32\textwidth]{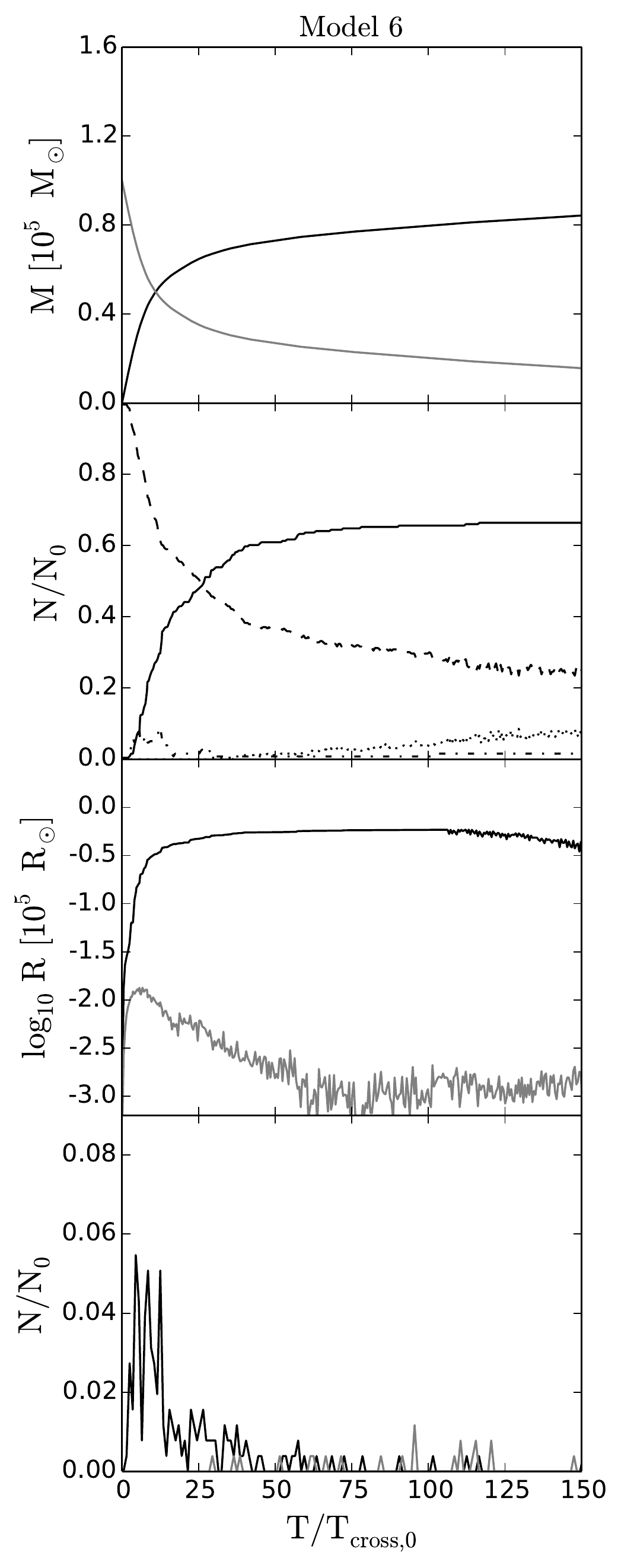} \\
\end{tabular}  
\caption{ Same as the previous figure but for the position dependent accretion models. }
\label{fig:time_evolutions2}
\end{figure*}

\begin{figure}
\centering
\begin{tabular}{c}
\includegraphics[height=0.38\textwidth,width=0.45\textwidth]{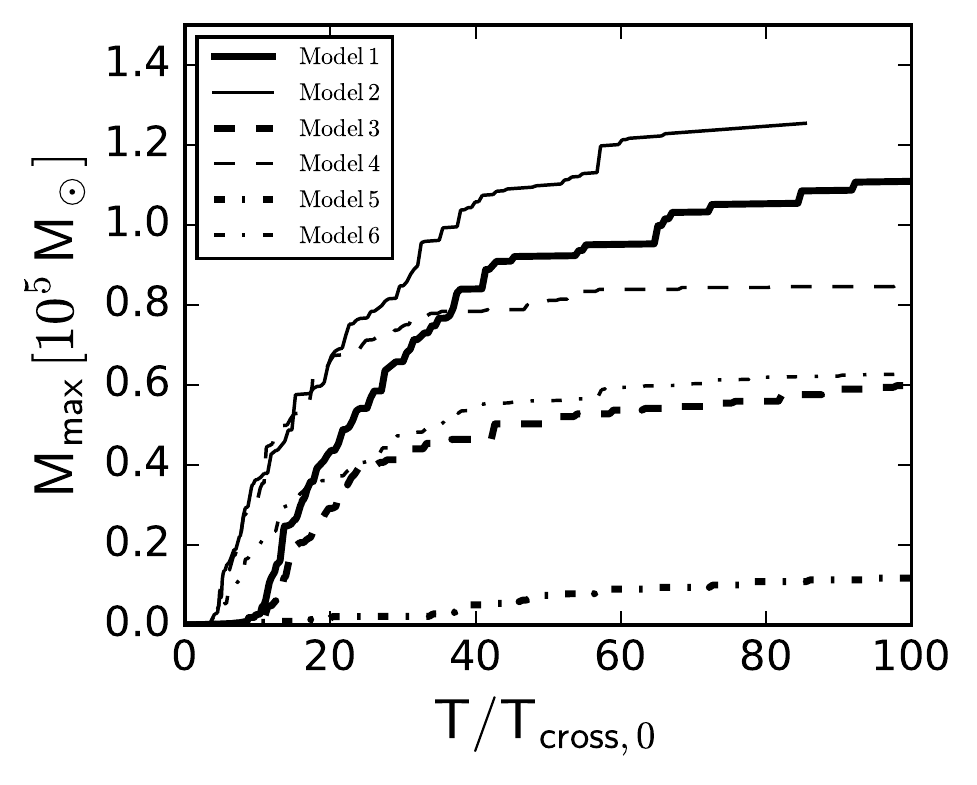} \\
\end{tabular}  
\caption{ Time evolution of the maximum mass in the system. We show the results for the six different accretion models and standard set of parameters. Except for Model 5, all models efficiently convert at least half of the initial gas mass into one single massive object.  }
\label{fig:individual_Mmax}
\end{figure}

\begin{figure}
\centering
\begin{tabular}{c}
\includegraphics[height=0.40\textwidth,width=0.45\textwidth]{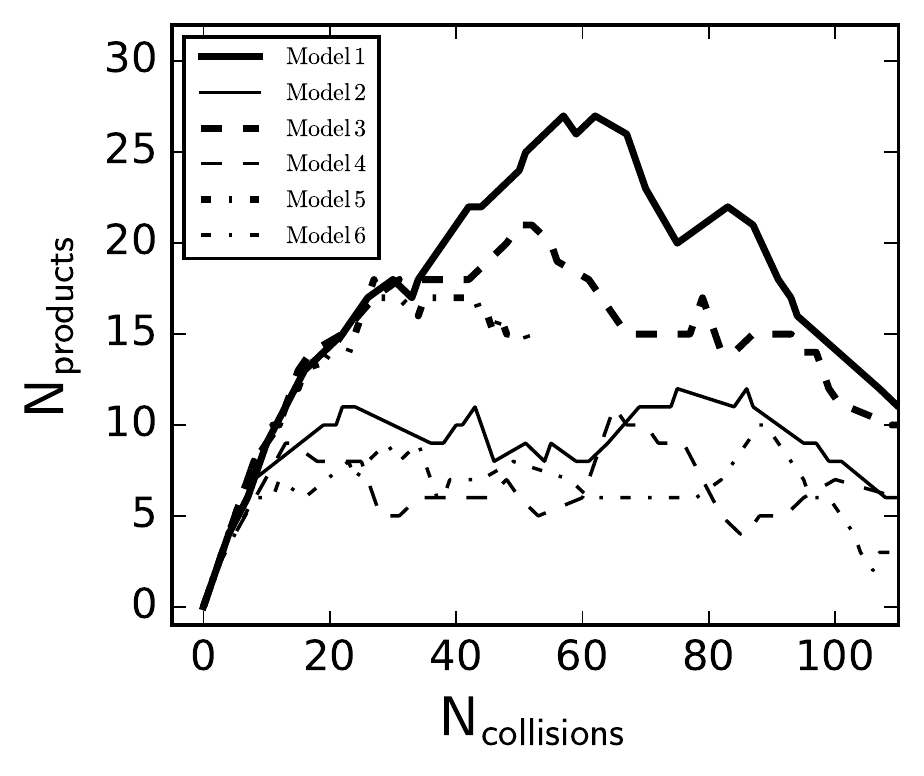} \\
\end{tabular}  
\caption{ Number of collision products present in the cluster as a function of number of collisions that have occurred. This is a measure of whether collisions occur with predominantly one massive object or between many different pairs of stars throughout the cluster. We clearly observe a difference between the radially independent (thick curves) and radially dependent (thin curves) accretion models. As is somewhat expected, the radially dependent accretion models produce a massive object in the center which is involved in most collisions.  }
\label{fig:individual_bully}
\end{figure}

Our simulations start with a cluster of single protostars, that is, there are no primordial binaries in the system.
In time collisions between protostars will occur and result in collision products. We distinguish between the most massive collision product and other less massive ones. Finally, strong  dynamical encounters between protostars can eject them from the cluster. We have thus defined four categories to which a protostar can belong: 

\begin{itemize}
\item Single protostar, i.e. not a collision product
\item Part of the most massive collision product
\item Part of a less massive collision product
\item Escaper
\end{itemize}

\noindent We note that binaries and higher order multiple systems form in our simulations, but we only categorize the individual stellar components. 
In Fig.~\ref{fig:relative_fractions} we present the time evolution of the fraction of protostars belonging to each of these four categories. 
We observe that the majority of protostars end up in the final most massive object, which can be understood intuitively as follows.  
As a result of the gas accretion, the protostellar radii are substantially increasing for two main reasons. First, the overall increase of the protostellar mass, which naturally provides larger radii. Second, the rather high accretion rates in the primordial environment, which lead to rather extended envelopes that may exceed several 1000 solar radii. As a consequence, the collision probability in this environment is highly enhanced once accretion is switched on, and once the effect of accretion onto the protostellar radii is being considered. In particular for the infinite gas reservoir models (1 and 2), we find that 80-90\% of the protostars end up in a single object. For finite but large gas reservoirs (as expected in the large atomic cooling halos), the fraction is somewhat reduced to 60-70\% in models 3, 4 and 6, but still quite significant. In model 5 the fraction reduces to about 10\% as due to the uniform accretion the central object in that cluster is less pronounced, and due to the time dependent accretion, the protostellar radii shrink again with decreasing gas reservoir. The latter represents an enhancement by roughly a factor of 10 compared to the corresponding gas-free models or models which assumed smaller protostellar radii \citep{2011MNRAS.413.1810B, Reinoso17}.
Comparing the radially independent accretion models (top row of Fig~\ref{fig:relative_fractions}) and the radially dependent models (bottom row), we observe a higher fraction of less massive collision products for the radially independent accretion models. Initially, this fraction is larger than the fraction of stars in the most massive collision product, implying that collisions tend to occur throughout the cluster between different pairs of stars. The radial independent models also produce a larger fraction of escaping protostars, implying that these systems produce stronger dynamical encounters. For the radially dependent accretion models most stars tend to immediately collide with the central protostar that is accreting the fastest. 

Next, we aim to better understand the time evolution of the collision rate of protostars. In Fig.~\ref{fig:time_evolutions} and \ref{fig:time_evolutions2} we present the collision rate (bottom panels) and correlate it with the total star and gas mass (top panels), the fraction of stars belonging to the four categories defined earlier (second panels) and the radius of the most massive protostar and the average radius of the remaining protostars (third panels). 

There is a small delay time for the collision rate to start to increase, which corresponds to the time it takes for the protostars to grow in mass and radius, and for the total stellar mass to become significant compared to the total gas mass. Then the collision rate increases rapidly and reaches a peak value. This rapid growth is fuelled by accretion which results in the protostars reaching larger sizes, and which makes the system stellar-mass dominated resulting in dynamical encounters and collisions. Values for the maximum and average collision rates can be found in Tab.~\ref{tab:simulation_overview2}, which together specify a range of typical collision rates in our simulations. We find that the maximum collision rate can be very high, i.e. a collisional fraction per crossing time up to about 8\% (see left panels in Fig.~\ref{fig:time_evolutions} and \ref{fig:time_evolutions2}), but that the average collision rate is more consistent with previous studies. For example, \citet[][Fig.~4]{2011MNRAS.413.1810B} measure a collisional fraction of 0.1-1$\%$, and this result is confirmed by \citet{Reinoso17} for gas free systems. 
\citet{2004Natur.428..724P} estimate the collision rate in the cluster MGG-11 to be 10-100 collisions within the first $3\,\rm{Myr}$, or since the crossing time is about $10^5$\,yr \citep{2003ApJ...596..240M}, $\sim 0.3-3$ collisions per crossing time}, which is comparable to the average rates in Tab.~\ref{tab:simulation_overview2}. 
Our results show that accretion-induced collisions in massive Pop.~III protostar clusters can increase the peak collision rate by an order of magnitude or even more. 

The saturation and eventual decrease of the collision rate has several causes. For the infinite gas reservoir models, the collision rate decreases again due to the sparsity of collision partners left in the system. For the finite gas reservoir models, the collision rate decreases due to a lack of gas, i.e. the stars do not expand anymore due to accretion. In this regime we observe that the fraction of escaping stars starts to increase, which implies strong dynamical encounters. After the initial evolution with a high accretion-induced collision rate, the system experiences a transition into a regime with a dynamics dominated collision rate, which is consequently much lower. For the time dependent accretion models the collision rate also decreases due to the shrinking of the protostars. We note however, that even though the collision rate decreases, that collisions that do occur can be quite massive if both collision partners are collision products themselves. 

In Fig.~\ref{fig:individual_Mmax} we plot the maximum mass in the system as a function of time. 
We observe that all models produce a very massive object with a mass of order $10^{4-5}\,\rm{M_\odot}$. 
The fastest growth in mass occurs between 5-20 crossing times, which is synchronous to the moment of highest collision rate (see Fig.~\ref{fig:time_evolutions} and \ref{fig:time_evolutions2} bottom panels). 
For Model 5, where accretion-induced collisions have the least effect, there is a longer delay time of about 30-50 crossing times, consistent with results from \citet[][Fig.~5]{Reinoso17}. 

The results presented so far show that the formation of a central massive object is the natural outcome, in the regime of high accretion rates and taking into account the large radii of accreting Pop.~III protostars. It is favourable to have a position dependent accretion rate, such that the formation of a massive object in the core is more likely, which fuels subsequent collisions with other stars. In Model 5 such a formation channel is missing. This plus the fact that the stars will eventually shrink due to the time-dependent accretion rate, causes the massive object to be an order of magnitude less massive. Model 5 comes closest to resembling the evolution of a gas free, equal mass Plummer sphere in which collisions usually do not become important until core collapse. 
This implies that for systems in which accretion-induced collisions do play a significant role, the formation of a massive object is sped up.  

Next, we want to further investigate how the collisions occur in the cluster: do they preferably happen with a dominant central object, or are there collisions between different pairs of objects occurring throughout the cluster? In Fig.~\ref{fig:individual_bully} we plot the number of collision products in the cluster as a function of the total number of collisions that have occurred. We observe that the position independent accretion models (black curves) can produce a factor 2-3 more collision products than the position dependent models (grey curves). This is somewhat expected considering that the position dependent accretion models rapidly produce a dominant mass in the core, which will have a much larger cross section for collisions. We note that the mass ratio between the two most massive objects in the system ranges from an order of magnitude for Model 5 up to two orders of magnitude for the other models.     

\subsection{Variation of input parameters}

\begin{figure*}
\centering
\begin{tabular}{cc}
\includegraphics[height=0.45\textwidth,width=0.45\textwidth]{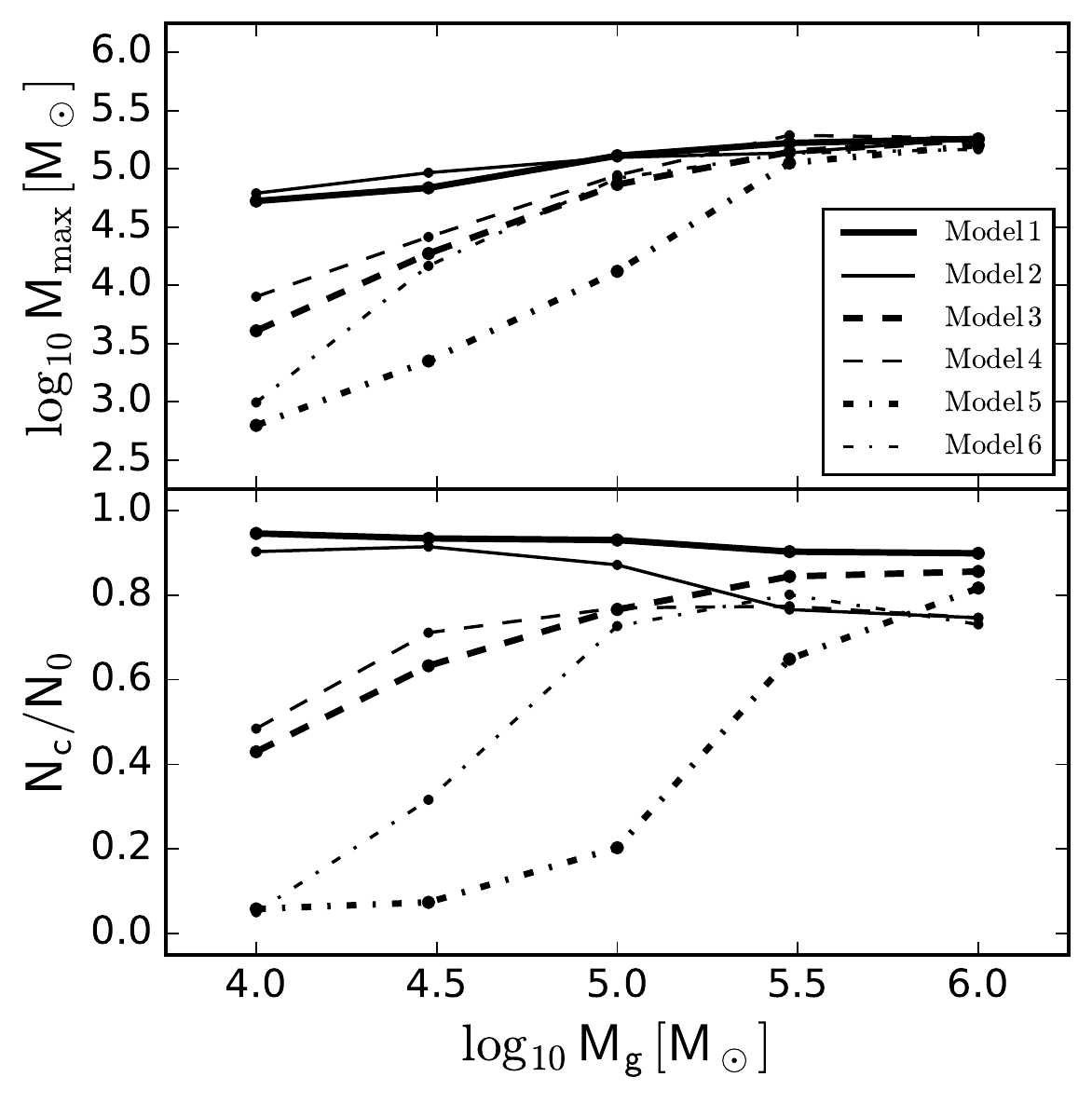} &
\includegraphics[height=0.45\textwidth,width=0.45\textwidth]{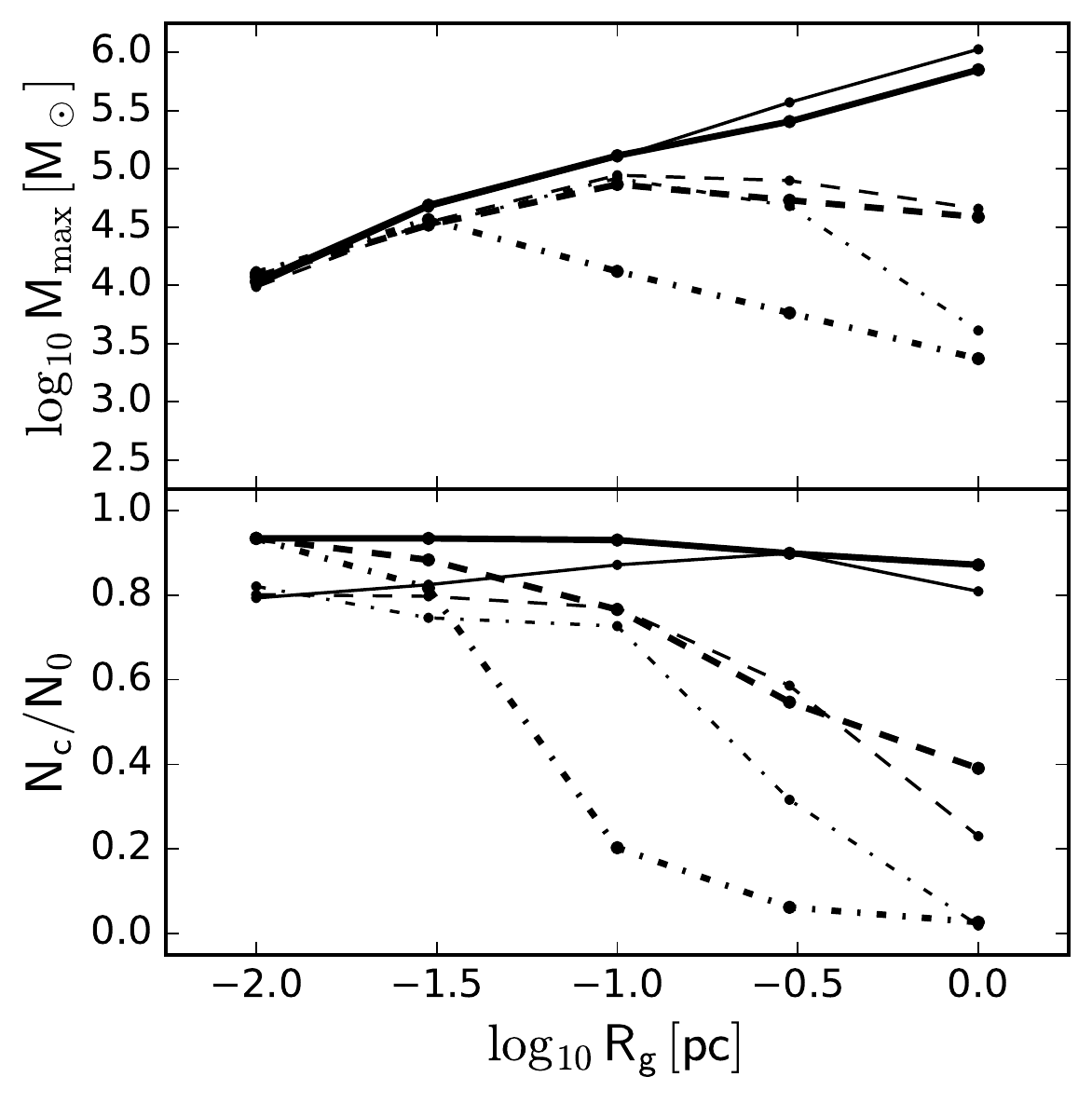} \\

\includegraphics[height=0.45\textwidth,width=0.45\textwidth]{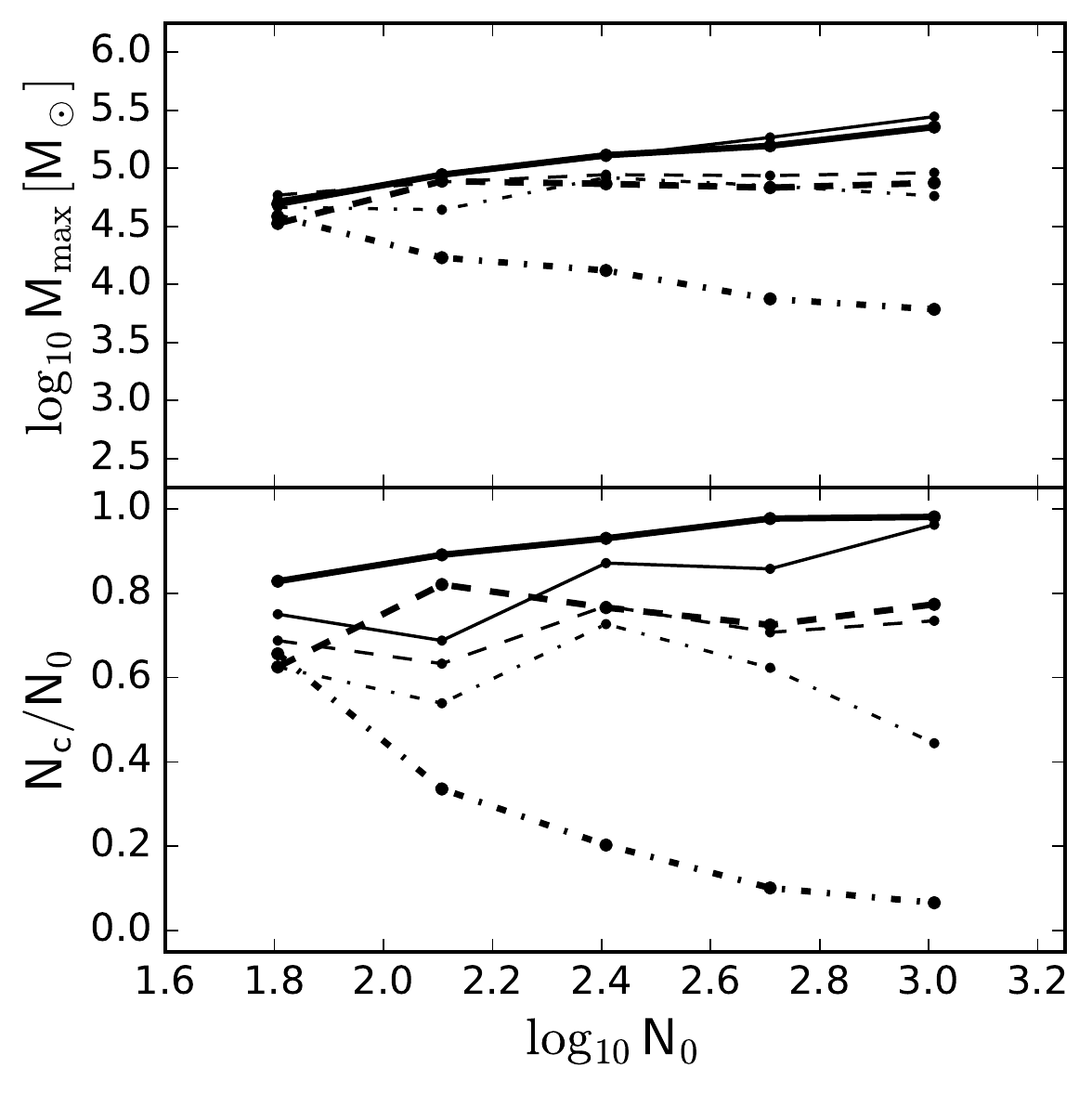} &
\includegraphics[height=0.45\textwidth,width=0.45\textwidth]{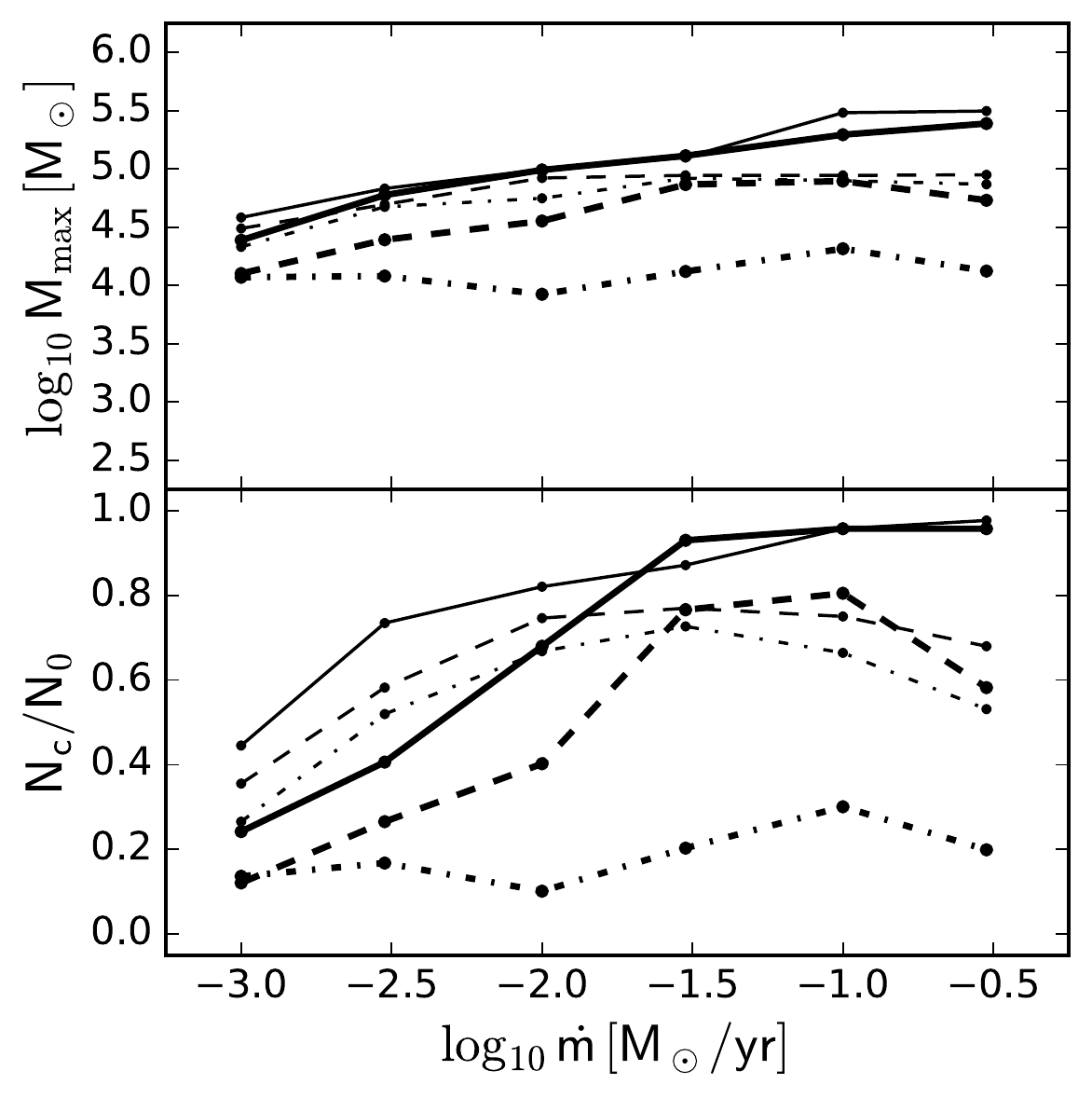} \\

\end{tabular}  
\caption{ Mass of the most massive object in the system at the end of the simulation, $M_{\rm{max}}$, and the total fraction of collisions, $N_c / N_0$, as a function of the model input parameters: initial gas cloud mass (top left), gas cloud radius (top right), initial number of protostars (bottom left) and initial average accretion rate per protostar (bottom right).  }
\label{fig:statistics}
\end{figure*}

\begin{table*}
\centering
\begin{tabular}{lrrrrrrrrrrrr}
\hline
Model & $M_{max,1}$ & $\delta M_{max,1}$ & $M_{max,2}$ & $\delta M_{max,2}$ & $N_{\rm{col,1}}$ & $\delta N_{\rm{col,1}}$ & $N_{\rm{col,2}}$ & $\delta N_{\rm{col,2}}$ & $R_{max,1}$ & $\delta R_{max,1}$ & $R_{max,2}$ & $\delta R_{max,2}$ \\
\hline 			
      & $\rm{M_\odot}$ & $\rm{M_\odot}$ & $\rm{M_\odot}$ & $\rm{M_\odot}$ & & & & & $\rm{t_{cross,0}^{-1}}$ & $\rm{t_{cross,0}^{-1}}$ & $\rm{t_{cross,0}^{-1}}$ & $\rm{t_{cross,0}^{-1}}$ \\
\hline 			
1 & 120529 & 13127 & 173130 & 13856 & 235.3 & 6.67 & 182.5 & 6.17 & 23.3 & 5.48 & 14.2 & 3.01 \\
2 & 139439 & 16503 & 160317 & 16480 & 228.6 & 5.30 & 225.4 & 3.60 & 28.9 & 4.12 & 25.7 & 6.60 \\
3 & 74613 & 3777 & 33329 & 4637 & 201.5 & 6.00 & 87.4 & 10.60 & 16.8 & 3.68 & 5.4 & 1.43 \\
4 & 84285 & 4581 & 83162 & 5962 & 189.1 & 13.74 & 181.1 & 15.88 & 21.7 & 3.47 & 19.6 & 4.58 \\
5 & 13381 & 1791 & 9238 & 1174 & 47.3 & 4.79 & 25.3 & 3.30 & 3.7 & 0.82 & 2.3 & 1.06 \\
6 & 71523 & 3729 & 56040 & 2352 & 183.2 & 5.71 & 121.7 & 9.44 & 18.8 & 4.08 & 12.8 & 2.20 \\
\hline
\end{tabular}
\caption{ Comparing two different mass-radius parametrizations: \citet{Hosokawa2012} (subscript 1) and \citet{2017arXiv170509301H} (subscript 2) (defined in Sec.~\ref{sec:mrmode}). We compare their influence on the formation of a massive object: its final mass $M_{\rm{max}}$, number of collisions $N_{\rm{col}}$ and maximum collision rate $R_{\rm{max}}$. We compare runs with the exact same initial realization of the cluster, and calculate the average and standard deviation over 10 simulations. Models 2 and 4 give consistent results, which are the position dependent accretion models where a dominant object is formed in the core. For the other models we observe a statistical difference due to the more conservative radii of \citet{2017arXiv170509301H}. This difference does not change the conclusion that a single massive object is formed due to a high collision rate.  }
\label{tab:mrmode_comparison}
\end{table*}

We perform an ensemble of numerical experiments with varying input parameters to determine the robustness of the result in the previous section, i.e. that a single massive object of mass $\sim 10^{4-5}\,\rm{M_\odot}$ is formed due to a high collision rate. We vary the initial total gas mass, gas cloud radius, number of protostars and average accretion rate per protostar. At the end of the simulation we determine the maximum mass in the system and the total number of collisions. The results are presented in Fig.~\ref{fig:statistics}. 

Starting with the total gas mass, we observe that the maximum mass increases with higher gas mass. This makes sense intuitively as more gas is available to accrete from, and the resulting protostars will be more massive and larger in size. This increases the collisional cross section, which is confirmed by the increasing number of collisions as a function of $M_g$. The trend seems to saturate at a gas mass of about $10^{5.5}\,\rm{M_\odot}$, and all the different models start to converge. In this regime, the protostars are able to reach such large radii that a runaway collision process is triggered irrespective of the detailed accretion model. 
At the low mass end of the diagram, we observe that the time dependent accretion models become inefficient, producing maximum masses $\le 10^3\,\rm{M_\odot}$. With little mass to accrete from, the protostars in the time-dependent models will generally remain small, resulting in a low collision rate, which is consistent with gas free star clusters in virial equilibrium.   

In the top right panel of Fig.~\ref{fig:statistics} we present the maximum mass and number of collisions with varying gas cloud radius. For the infinite gas reservoir models, we observe an increasing trend. For a larger gas cloud radius, the protostars need to accrete more gas before they reach the required radius for collisions to occur. As a result the collision products will be more massive. 
For the finite gas reservoir models there is a peak at $R_{\rm{g}} \sim 10^{-1.5}-10^{-1}\,\rm{pc}$ after which the maximum mass decreases again. Intuitively, we can relate this decrease to the decrease in the number density, which is proportional to the collision rate. This is confirmed by the decrease of the number of collisions with $R_g$.  
For the densest configurations, i.e. the smallest gas cloud radii, we again observe a convergence of the different accretion models. With decreasing $R_g$ the ratio of the protostellar radii to $R_g$ increases, resulting in an increased collision rate, which is confirmed by the data. The trend that $M_{\rm{max}}$ increases with $R_g$, for the densest configurations, is explained by the fact that the protostars have to accrete for a longer time to reach a significant radius relative to $R_g$, thus resulting in more massive objects. The initial increase and subsequent decrease of $M_{\rm{max}}$ with increasing $R_g$ is thus explained by the increase of the accretion time scale and the decrease of the number density.    

In the bottom left panel of Fig.~\ref{fig:statistics} we vary the number of initial protostars in the cluster. Intuitively, we would expect that with a higher number of stars, the collision rate will increase. However, with fewer stars in the cluster, they will be able to accrete more and thus increase their collisional cross section. For the infinite gas reservoir models we again observe a steady increase of the maximum mass with $N$. The finite models produce massive objects between $10^{4.5-5}\,\rm{M_\odot}$ irrespective of the variation of $N$ from 64 to 1024. The total collision fraction also remains roughly constant between 0.6 and 0.8. Therefore the increasing number density and decreasing cross section with $N$ tend to cancel out. Only for Model 5 we observe a decrease of $M_{\rm{max}}$ and $N_c/N_0$ with $N$. The extra effect of the time dependent accretion rate and subsequent shrinking of the protostellar radii, causes the decreasing collisional cross section to dominate. For the time dependent accretion models it is therefore favourable to have fewer stars in the system in order to produce a massive object.      

In the bottom right panel of Fig.~\ref{fig:statistics} we plot the maximum mass and total collision fraction as a function of average accretion rate per protostar. We observe an increasing collision fraction as a function of $\dot{m}$. This can be explained by the mass-radius tracks: a higher $\dot{m}$ produces larger protostellar radii. For the largest values of $\dot{m}$ however, the variation in the protostellar radii saturates (see Fig.~\ref{fig:lookup}), and we observe a flattening of the total collision fraction.  
The maximum mass depends weakly on $\dot{m}$, showing a relatively mild increase. For our standard set of parameter values defined Sec.~\ref{sec:exp_setup}, we conclude that for $\dot{m} \ge 10^{-3}\,\rm{M}_\odot / \rm{yr}$ a massive object of $M \ge 10^4\,\rm{M}_\odot$ is formed.   

Finally, we measure the dependence of the formation of a massive object to a different underlying mass-radius parametrization. We compare our parametrization based on  \citet{Hosokawa2012} to that of \citet{2017arXiv170509301H}. We use the standard set of parameters defined in Sec.~\ref{sec:exp_setup} and create 10 initial realizations of the cluster. We integrate these initial conditions using both mass-radius parametrizations and calculate the average and standard deviation of the final maximum mass, number of collisions and maximum collision rate (see Tab.~\ref{tab:mrmode_comparison}). For models 2 and 4 we find the results to be mutually consistent. In these radial dependent accretion models a dominant central object is formed in the core. Its rapid growth causes many collisions to occur irrespective of the detailed mass-radius parametrization. For models 3, 5 and 6 we observe that the models based on \citet{2017arXiv170509301H} produce somewhat fewer collisions and less massive objects. These models are sensitive to the more conservative radii of the \citet{2017arXiv170509301H} parametrization because the accretion model is radially independent (so no dominant object in the core due to accretion) and/or because the accretion model is time dependent, in which protostars shrink to small sizes and differences in the radii will have a large effect on the dynamics dominated collision rate.  
For Model 1 we observe the counter-intuitive result that the parametrization by \citet{2017arXiv170509301H} produces fewer collisions but a more massive object. This can be explained by the difference in time at which the stopping condition was fulfilled. The collisions rate is lower and the collisions therefore more spread out in time. In this infinite gas reservoir model this gives the protostars more time to accrete mass.

\section{Conclusion}\label{sec:conclusion}

We present the first numerical study of the formation of massive black hole seeds through the formation channel of accretion and collisions in a dense Population~III (Pop.~III) protostellar cluster. We take into account accurate mass-radius parametrizations for accreting Pop.~III protostars, which are based on detailed stellar evolution calculations performed by \citet{Hosokawa2012} and \citet{2017arXiv170509301H}. These studies have shown that for high accretion rates ($\dot{m} \ge 10^{-3}\,\rm{M}_\odot/ \rm{yr}$), the protostellar radii can become of order $10^{2-4}\,\rm{R}_\odot$, thus significantly increasing the collisional cross section. Using the \texttt{AMUSE} simulation framework we perform a series of multi-physics simulations, including $N$-body dynamics, an analytical gas potential, six different accretion models, mass-radius parametrizations and stellar collisions.  

In Tab.~\ref{tab:simulation_overview} and \ref{tab:simulation_overview2} we present an overview of the simulation input parameters and outcome statistics. 
Our most conservative model (model 5), which most closely resembles the dynamics of ordinary Plummer spheres, confirms the fractional collision rate of 0.1 - 1\% per crossing time found in previous studies.
Depending on the values of the input parameters, about 10\% of the initial protostellar population collides into a single massive object of order $10^4\,\rm{M}_\odot$ on a time scale of 50 crossing times or longer. 

In our remaining accretion models, the effects of accretion-induced collisions are more prominent for two distinct but related reasons. 
First, in case of a more extensive gas reservoir the protostars will have more time to accrete gas and as a result be able to reach larger radii. 
Second, massive objects form in the center of the cluster where gas volume densities, and thus the accretion rates and collision rates, are highest.  
Our results show that the maximum number of collisions per crossing time can be increased up to 1-10\% of the initial protostellar population. This increased collision rate leads to the formation of a single massive object with a mass of order $M = 10^{4-5}\,\rm{M}_\odot$. The increase of the collision rate is fuelled by the high accretion rates of the Pop.~III protostars and the resulting large radii. After the maximum collision rate is reached, the collision rate decreases again, either due to the sparsity of collision partners left in the system, or due to the lack of gas, which truncates the growth of the protostars and allows for dynamical ejections from the cluster. In Fig.~\ref{fig:time_evolutions}, \ref{fig:time_evolutions2} and \ref{fig:individual_Mmax} we also demonstrate that the formation of a massive object can occur within 20 crossing times, which is about twice as fast compared to gas free models (e.g. Reinoso et al. 2017). 

We have varied the initial gas cloud mass and gas cloud radius, and confirm the intuitive result that most collisions occur in the densest systems (see Fig.~\ref{fig:statistics} top two panels). Also, the final mass of the seed black hole increases with higher gas cloud mass. Our calculations based on models with a limited gas reservoir, show a characteristic gas cloud radius of $R_g = 10^{-1.5}-10^{-1}\,\rm{pc}$, at which the final mass of the most massive object is a maximum, i.e. $M \sim 10^{4.5-5}\,\rm{M}_\odot$. For smaller gas cloud radii, the system becomes more dense, which triggers a runaway collision between the protostars at earlier times. The less time there is for the protostars to accrete gas, the less massive will be the collision products. In the opposite regime of larger gas cloud radii, a runaway collision might not occur as the protostars cannot reach the required radii. These systems eventually become stellar-mass dominated, and since the collision rate is proportional to the number density, we would expect fewer collisions in larger systems. The final mass of the seed black hole depends mildly on the initial number of protostars (ranging from $N=64-1024$) and average accretion rate (ranging from $\dot{m} = 0.001 - 0.3\,\rm{M}_\odot / \rm{yr}$), consistently producing massive objects of order $M=10^4\,\rm{M}_\odot$ or higher.   

The mass-radius evolution of accreting Pop.~III protostars is very uncertain. In order to estimate the associated uncertainty and the robustness of the results mentioned above, we compare two mass-radius parametrizations: one based on \citet{Hosokawa2012} and a more conservative model based one \citet{2017arXiv170509301H}. Our calculations confirm that a parametrization with more conservative radii results in lower collision rates. However, the two different mass-radius parametrizations, each based on detailed stellar evolution calculations, produce no strong differences in the final mass of the seed black hole. 

We conclude that the mechanism of accretion-induced collisions in dense, Pop.~III protostellar systems is a viable mechanism for explaining the formation of the first massive black hole seeds. Therefore this investigation warrants follow up studies, which improve on the realism of the detailed implementation and coupling of the different physics ingredients.     

\section{Acknowledgements}

TB acknowledges support from Funda\c{c}\~ ao para a Ci\^ encia e a Tecnologia (grant SFRH/BPD/122325/2016), and support from Center for Research \& Development in Mathematics and Applications (CIDMA) (strategic project UID/MAT/04106/2013), and from ENGAGE SKA, POCI-01-0145-FEDER-022217, funded by COMPETE 2020 and FCT, Portugal.
DRGS and BR thank for funding through Fondecyt regular (project code 1161247). financial support. 
BR thanks Conicyt for financial support (CONICYT-PFCHA/Mag\'isterNacional/2017-22171385). 
DRGS, MF and AMS acknowledge funding through the ''Concurso Proyectos Internacionales de Investigaci\'on, Convocatoria 2015'' (project code PII20150171) and the BASAL Centro de Astrof\'isica y Tecnolog\'ias Afines (CATA) PFB-06/2007. 
DRGS further is grateful for funding via ALMA-Conicyt (project code 31160001), Quimal (project code QUIMAL170001), and the Anillo program ``Formation and Growth of Supermassive Black Holes'' (project code ACT172033). 
RSK acknowledges financial support from the Deutsche Forschungsgemeinschaft via SFB 881, ``The Milky Way System'' (sub-projects B1, B2 and B8) and SPP 1573, ``Physics of the Interstellar Medium''. 
LH and RSK were supported by the European Research Council under the European Community's Seventh Framework Programme (FP7/2007 - 2013) via the ERC Advanced Grant `STARLIGHT: Formation of the First Stars' (project number 339177).

\bibliographystyle{mn2e}      
\bibliography{massive_stars_arxiv}      

\appendix

\section{Analytical form of the mass-radius relation}\label{massradius}

The analytical power-law parametrization of the mass-radius relation from the models of \citet{Hosokawa2012} and \citet{2017arXiv170509301H} is given in Tables~\ref{tab:param} and \ref{tab:param2}, respectively.

\begin{table*}
\centering
\begin{tabular}{c|c|c}
\hline
$\dot{\rm{m}}$ & m & R \\
\hline
$\rm{M}_\odot\,\rm{yr}^{-1}$ & $\rm{M}_\odot$ & $\rm{R}_\odot$ \\
\hline
$\geq \rm{1}$ & $< \rm{150}$    & $\rm{250} \left( {\rm{m} \over \rm{1}\,\rm{M}_\odot} \right)^{\rm{0.4}}$ \\
              & $< \rm{450}$    & $\rm{1855} \left( {\rm{m} \over \rm{150}\,\rm{M}_\odot} \right)^{\rm{0.8}}$ \\
              & $\geq \rm{450}$ & $\rm{4468} \left( {\rm{m} \over \rm{450}\,\rm{M}_\odot} \right)^{\rm{0.5}}$ \\
\hline
$ \rm{0.3} \leq \dot{m} < \rm{1}$ & $< \rm{80}$    & $\rm{190} \left( {\rm{m} \over \rm{1}\,\rm{M}_\odot} \right)^{\rm{0.4}}$ \\
                                  & $< \rm{140}$    & $\rm{1096} \left( {\rm{m} \over \rm{80}\,\rm{M}_\odot} \right)^{\rm{1.5}}$ \\
                                  & $\geq \rm{140}$ & $\rm{2538} \left( {\rm{m} \over \rm{140}\,\rm{M}_\odot} \right)^{\rm{0.5}}$ \\    
\hline
$ \rm{0.1} \leq \dot{m} < \rm{0.3}$ & $< \rm{40}$    & $\rm{140} \left( {\rm{m} \over \rm{1}\,\rm{M}_\odot} \right)^{\rm{0.4}}$ \\
                                  & $< \rm{90}$    & $\rm{612} \left( {\rm{m} \over \rm{40}\,\rm{M}_\odot} \right)^{\rm{1.5}}$ \\
                                  & $\geq \rm{90}$ & $\rm{2066} \left( {\rm{m} \over \rm{90}\,\rm{M}_\odot} \right)^{\rm{0.5}}$ \\                                             
\hline
$ \rm{0.06} \leq \dot{m} < \rm{0.1}$ & $< \rm{25}$    & $\rm{110} \left( {\rm{m} \over \rm{1}\,\rm{M}_\odot} \right)^{\rm{0.4}}$ \\
                                  & $< \rm{70}$    & $\rm{399} \left( {\rm{m} \over \rm{25}\,\rm{M}_\odot} \right)^{\rm{1.5}}$ \\
                                  & $\geq \rm{70}$ & $\rm{1868} \left( {\rm{m} \over \rm{70}\,\rm{M}_\odot} \right)^{\rm{0.5}}$ \\                                             
\hline
$ \rm{0.03} \leq \dot{m} < \rm{0.06}$ & $< \rm{20}$    & $\rm{90} \left( {\rm{m} \over \rm{1}\,\rm{M}_\odot} \right)^{\rm{0.4}}$ \\
                                  & $< \rm{70}$    & $\rm{298} \left( {\rm{m} \over \rm{20}\,\rm{M}_\odot} \right)^{\rm{1.5}}$ \\
                                  & $\geq \rm{70}$ & $\rm{1953} \left( {\rm{m} \over \rm{70}\,\rm{M}_\odot} \right)^{\rm{0.5}}$ \\                                             
\hline
$ \rm{0.006} \leq \dot{m} < \rm{0.03}$ & $< \rm{20}$    & $\rm{50} \left( {\rm{m} \over \rm{1}\,\rm{M}_\odot} \right)^{\rm{0.4}}$ \\
                                  & $< \rm{35}$    & $\rm{166} \left( {\rm{m} \over \rm{20}\,\rm{M}_\odot} \right)^{\rm{-1.5}}$ \\
                                  & $\geq \rm{35}$ & $\rm{72} \left( {\rm{m} \over \rm{35}\,\rm{M}_\odot} \right)^{\rm{0.5}}$ \\                                             
\hline
$ \dot{m} < \rm{0.006}$ & $< \rm{9}$    & $\rm{25} \left( {\rm{m} \over \rm{1}\,\rm{M}_\odot} \right)^{\rm{0.4}}$ \\
                                  & $< \rm{50}$    & $\rm{60} \left( {\rm{m} \over \rm{9}\,\rm{M}_\odot} \right)^{\rm{-1.5}}$ \\
                                  & $\geq \rm{50}$ & $\rm{5} \left( {\rm{m} \over \rm{50}\,\rm{M}_\odot} \right)^{\rm{0.5}}$ \\                                             
\hline
\end{tabular}
\caption{ Simplified parametrization of the mass-radius relation of accreting Pop.~III protostars based on  \citet[][Fig.~5]{Hosokawa2012}.  }
\label{tab:param}
\end{table*}

\begin{table*}
\centering
\begin{tabular}{c|c|c}
\hline
$\dot{\rm{m}}$ & m & R \\
\hline
$\rm{M}_\odot\,\rm{yr}^{-1}$ & $\rm{M}_\odot$ & $\rm{R}_\odot$ \\
\hline
$\geq \rm{10}$ & $< \rm{10}$    & $\rm{200} \left( {\rm{m} \over \rm{1}\,\rm{M}_\odot} \right)^{\rm{0.0}}$ \\
              & $< \rm{50}$    & $\rm{200} \left( {\rm{m} \over \rm{10}\,\rm{M}_\odot} \right)^{\rm{-0.4}}$ \\
              & $< \rm{60}$ & $\rm{105} \left( {\rm{m} \over \rm{50}\,\rm{M}_\odot} \right)^{\rm{13}}$ \\
              & $\geq \rm{60}$ & $\rm{1124} \left( {\rm{m} \over \rm{60}\,\rm{M}_\odot} \right)^{\rm{0.5}}$ \\               
\hline
$ \rm{1} \leq \dot{m} < \rm{10}$ & $< \rm{10}$    & $\rm{200} \left( {\rm{m} \over \rm{1}\,\rm{M}_\odot} \right)^{\rm{0.0}}$ \\
              & $< \rm{35}$    & $\rm{200} \left( {\rm{m} \over \rm{10}\,\rm{M}_\odot} \right)^{\rm{-0.6}}$ \\
              & $< \rm{42}$ & $\rm{94} \left( {\rm{m} \over \rm{35}\,\rm{M}_\odot} \right)^{\rm{13}}$ \\
              & $\geq \rm{42}$ & $\rm{1009} \left( {\rm{m} \over \rm{42}\,\rm{M}_\odot} \right)^{\rm{0.5}}$ \\     
\hline
$ \rm{0.1} \leq \dot{m} < \rm{1}$ & $< \rm{10}$    & $\rm{200} \left( {\rm{m} \over \rm{1}\,\rm{M}_\odot} \right)^{\rm{0.0}}$ \\
              & $< \rm{25}$    & $\rm{200} \left( {\rm{m} \over \rm{10}\,\rm{M}_\odot} \right)^{\rm{-0.6}}$ \\
              & $< \rm{29.5}$ & $\rm{115} \left( {\rm{m} \over \rm{25}\,\rm{M}_\odot} \right)^{\rm{13}}$ \\
              & $\geq \rm{29.5}$ & $\rm{993} \left( {\rm{m} \over \rm{29.5}\,\rm{M}_\odot} \right)^{\rm{0.5}}$ \\                                             
\hline
$ \rm{0.01} \leq \dot{m} < \rm{0.1}$ & $< \rm{30}$    & $\rm{200} \left( {\rm{m} \over \rm{1}\,\rm{M}_\odot} \right)^{\rm{0.0}}$ \\
                                  & $< \rm{70}$    & $\rm{200} \left( {\rm{m} \over \rm{30}\,\rm{M}_\odot} \right)^{\rm{-2.5}}$ \\
                                  & $\geq \rm{70}$    & $\rm{24} \left( {\rm{m} \over \rm{70}\,\rm{M}_\odot} \right)^{\rm{0.5}}$ \\                                             
\hline
$ \dot{m} < \rm{0.01}$ & $< \rm{10}$    & $\rm{200} \left( {\rm{m} \over \rm{1}\,\rm{M}_\odot} \right)^{\rm{0.0}}$ \\
                                  & $< \rm{50}$    & $\rm{200} \left( {\rm{m} \over \rm{10}\,\rm{M}_\odot} \right)^{\rm{-2.5}}$ \\
                                  & $\geq \rm{50}$ & $\rm{3.6} \left( {\rm{m} \over \rm{50}\,\rm{M}_\odot} \right)^{\rm{0.5}}$ \\                                             
\hline
\end{tabular}
\caption{ Simplified parametrization of the mass-radius relation of accreting Pop.~III protostars based on \citet[][Fig.~2]{2017arXiv170509301H}.  }
\label{tab:param2}
\end{table*}

\section{Validation of the numerical method}\label{app:BB}

\begin{figure}
\centering
\begin{tabular}{cc}
\includegraphics[height=0.32\textwidth,width=0.38\textwidth]{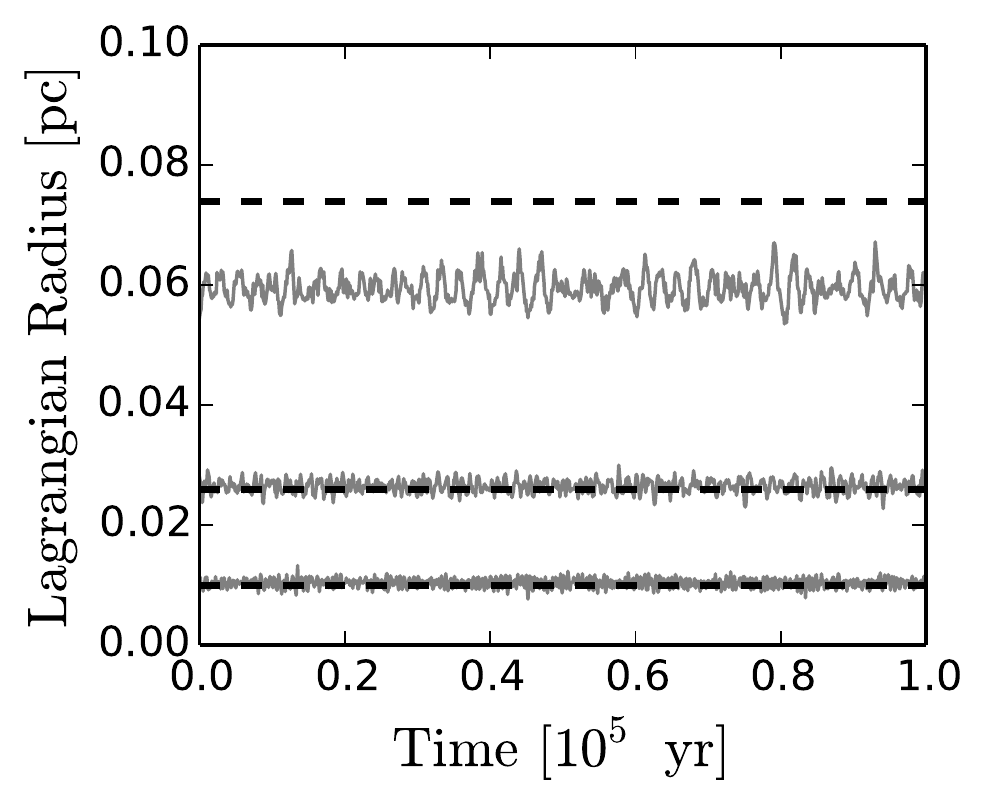} \\
\end{tabular}  
\caption{ Validation of the initial conditions and dynamics. We present the 10, 50 and 90$\%$ Lagrangian radius of a Plummer sphere (see text for the Plummer parameters), calculated analytically (dashed line, for an untruncated Plummer) and from our numerical model (solid line, truncated at 5 Plummer radii). We confirm that our cluster is initially stable and will only change its structure in the presence of accretion and collisions.  }
\label{fig:validation_dynamics}
\end{figure}

\begin{figure*}
\centering
\begin{tabular}{ccc}
\includegraphics[height=0.26\textwidth,width=0.3\textwidth]{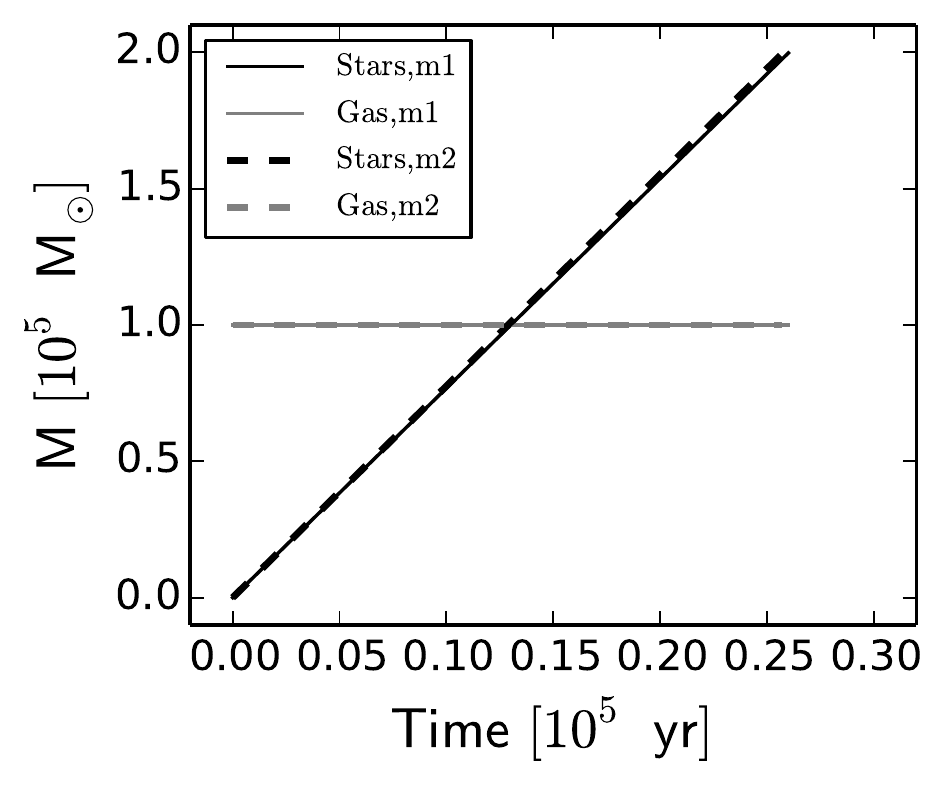} &
\includegraphics[height=0.26\textwidth,width=0.3\textwidth]{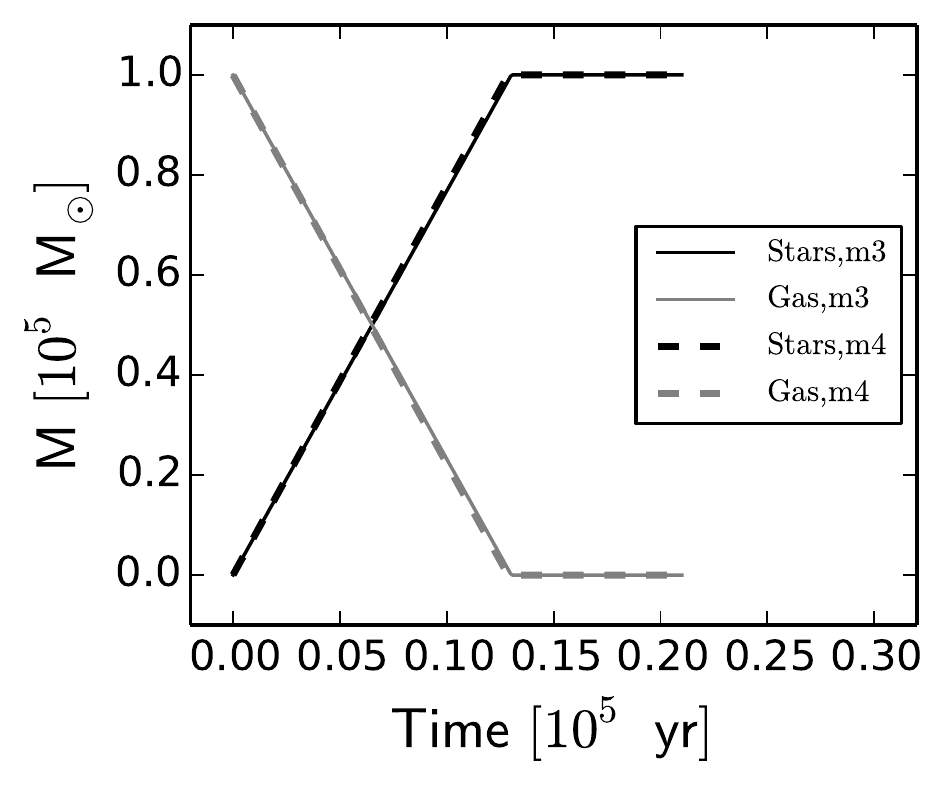} &
\includegraphics[height=0.26\textwidth,width=0.3\textwidth]{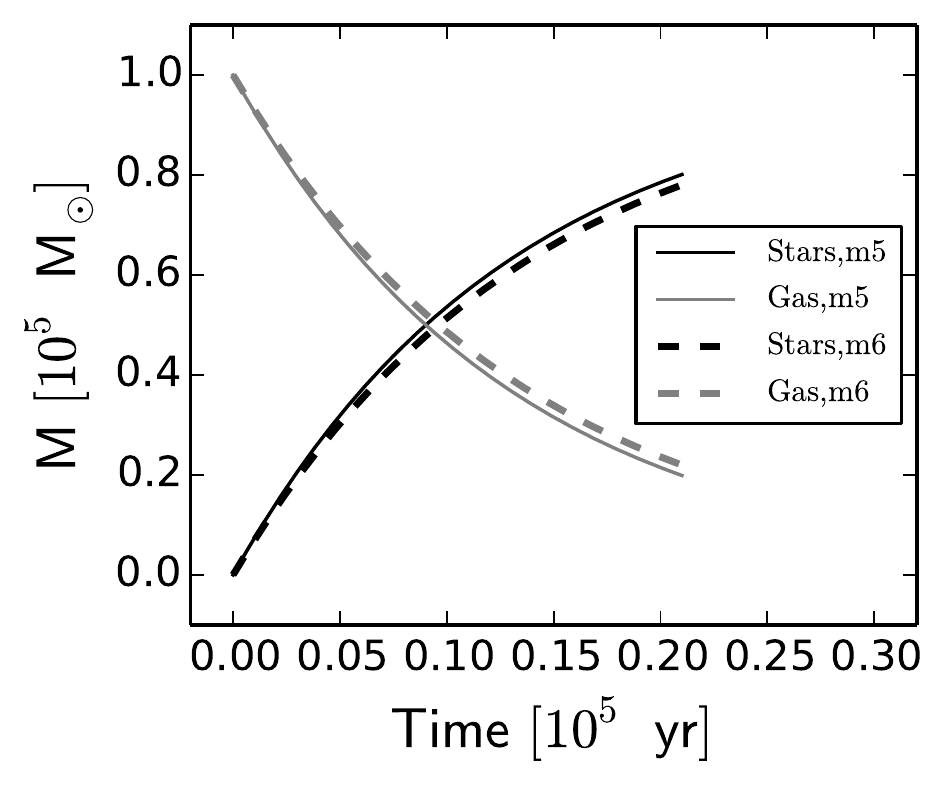} \\

\includegraphics[height=0.26\textwidth,width=0.3\textwidth]{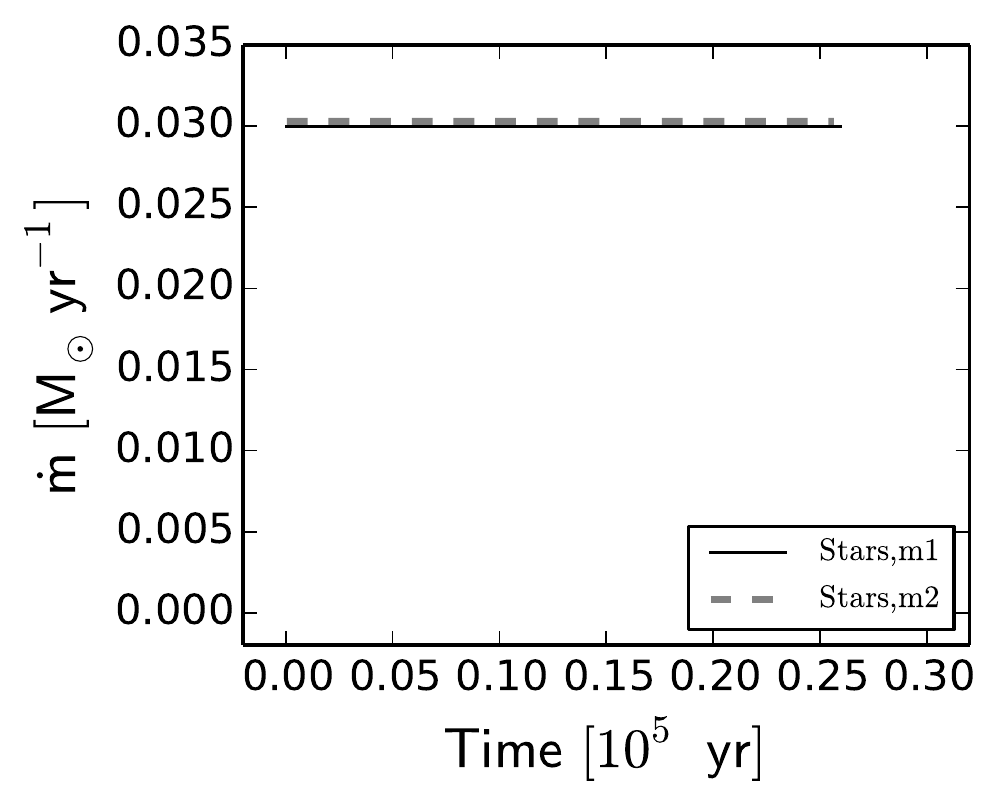} &
\includegraphics[height=0.26\textwidth,width=0.3\textwidth]{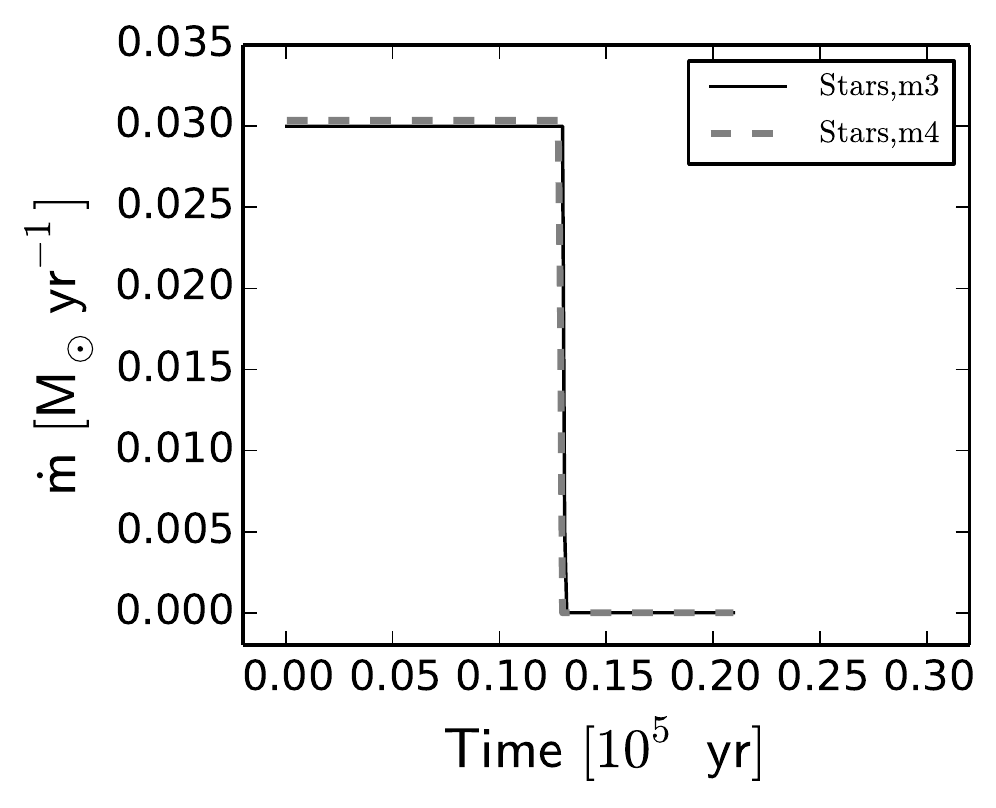} &
\includegraphics[height=0.26\textwidth,width=0.3\textwidth]{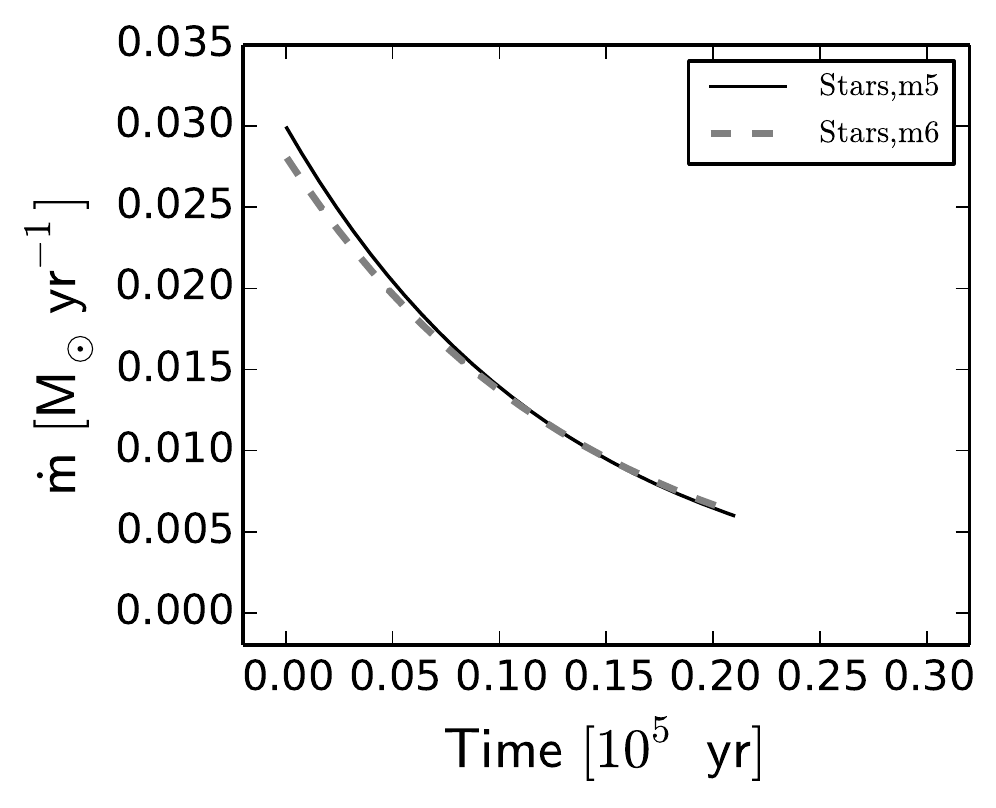} \\

\end{tabular}  
\caption{ Illustration and validation of the gas accretion models. In the top row we show the total mass evolution of the stars and gas. We confirm that models 1 and 2 are consistent (top left panel), and similar for models 3 and 4 (top middle panel) and models 5 and 6 (top right panel). In the bottom row we present the average accretion rates per star, i.e. the time derivative of the total stellar mass in the top row divided by the total number of stars.  }
\label{fig:validation_accretion}
\end{figure*}

\begin{figure}
\centering
\begin{tabular}{c}
\includegraphics[height=0.35\textwidth,width=0.41\textwidth]{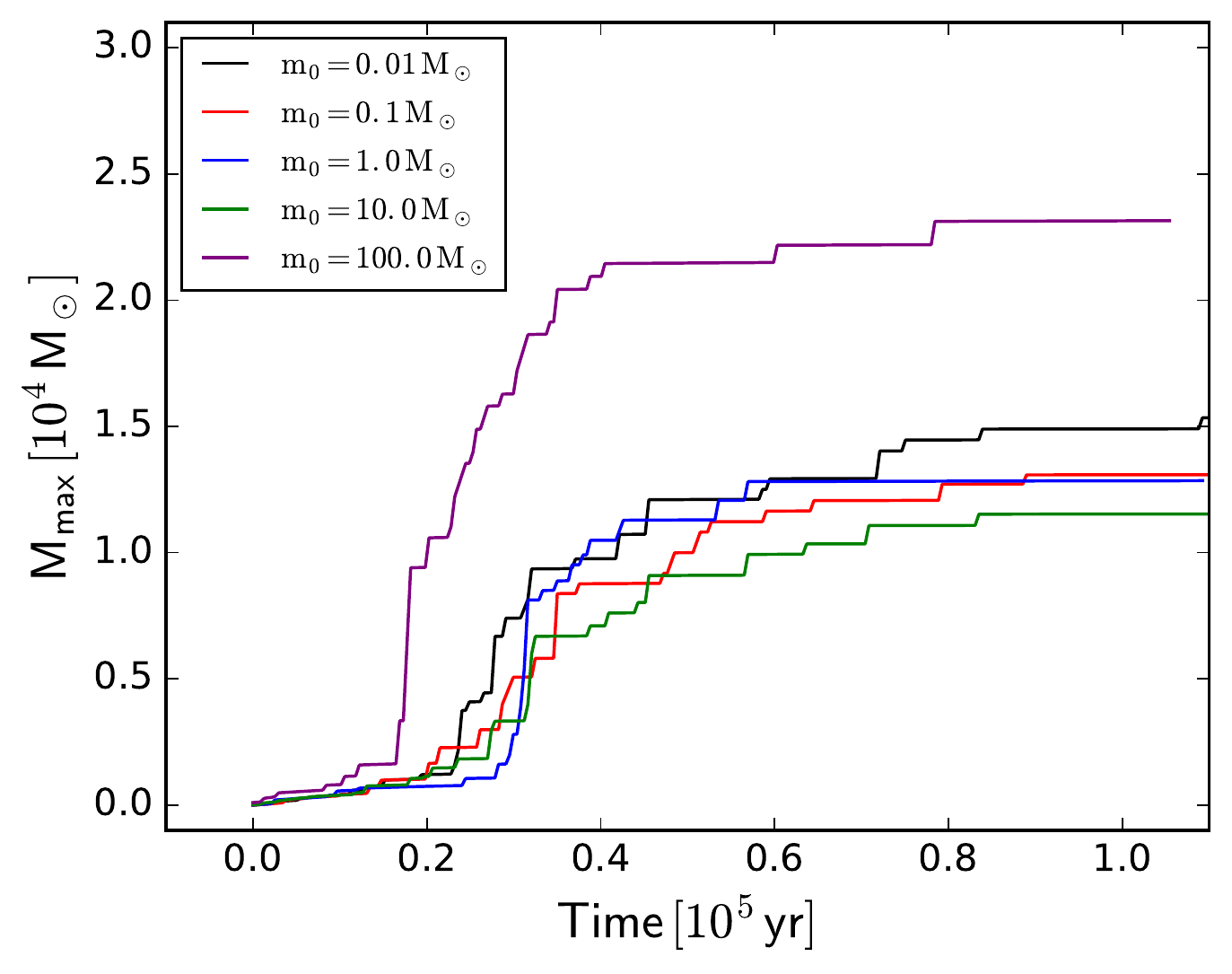} \\
\end{tabular}  
\caption{ { Time evolution of the maximum mass in the system, for our most conservative accretion model (Model 5) with standard parameters (see Sec.~3). We vary the initial masses of the protostars and find that as long as the system starts out as a gas-dominated system, that is with $m_0 \ll 390 \rm{M}_\odot$, the final maximum mass does not change much. For $m_0 = 100\,M_\odot$ the overall potential starts to become strongly influenced by the stars, and mass growth gets modified by Bondi-Hoyle-Littleton accretion. }  }
\label{fig:validation_m0}
\end{figure}

We present here additional plots for the validation experiments discussed in section~\ref{app:B}. In Fig.~\ref{fig:validation_dynamics} we confirm that in the absence of accretion, the 10, 50 and 90\% Lagrangian radii follow those of a Plummer sphere, with a slight discrepancy in the 90\% radius due to the truncation that we have introduced. In Fig.~\ref{fig:validation_accretion} we present the time evolution of the total star and gas mass (top row) and the time evolution of the average accretion rate (bottom row) in a setup where the $N$-body integrator has been switched off. The figure follows the expected behavior of the accretion models. { Finally, we show in Fig.~\ref{fig:validation_m0} that the specific choice of the initial protostellar mass does not have a large influence on the time evolution of the most massive object in the system. 
The initial total stellar mass is given by $M_s = N m_0$, which becomes comparable to the initial gas cloud mass, $M_g = 10^5\,\rm{M}_\odot$, and $N=256$, if $m_0 = 390.625\,\rm{M}_\odot$. When $m_0 \ll 390 \rm{M}_\odot$, the total mass (gas + stars) remains close to $\sim 10^5\,\rm{M}_\odot$, and with a fixed collisional fraction of about 10 percent, produces a maximum mass of about $10^4 \rm{M}_\odot$. When $m_0=100\,\rm{M}_\odot$, the initial stellar mass $M_s=2.56 \times 10^4 \rm{M}_\odot$, and so we have effectively increased the initial total mass by a quarter, which also allows for the formation of a more massive object. However, such a system would not start out as a gas-dominated system. }

\end{document}